\documentstyle[12pt]{article}
\def\beq{\begin{equation}}
\def\eeq{\end{equation}}

\def\beqn{\begin{eqnarray}}
\def\eeqn{\end{eqnarray}}
\relax
\catcode`\@=11
\@addtoreset{equation}{section}

\def\@normalsize{\@setsize\normalsize{15pt}\xiipt\@xiipt
\abovedisplayskip 14pt plus3pt minus3pt%
\belowdisplayskip \abovedisplayskip
\abovedisplayshortskip  \z@ plus3pt%
\belowdisplayshortskip  7pt plus3.5pt minus0pt}

\def\small{\@setsize\small{13.6pt}\xipt\@xipt
\abovedisplayskip 13pt plus3pt minus3pt%
\belowdisplayskip \abovedisplayskip
\abovedisplayshortskip  \z@ plus3pt%
\belowdisplayshortskip  7pt plus3.5pt minus0pt

\def\@listi{\parsep 4.5pt plus 2pt minus 1pt
            \itemsep \parsep
            \topsep 9pt plus 3pt minus 3pt}}

\def\underline#1{\relax\ifmmode\@@underline#1\else
	$\@@underline{\hbox{#1}}$\relax\fi}
\@twosidetrue

\catcode`@=12

\catcode`\@=11

\catcode`\@=12

\relax

\def\figcap{\section*{Figure Captions\markboth
	{FIGURECAPTIONS}{FIGURECAPTIONS}}\list
	{Fig. \arabic{enumi}:\hfill}{\settowidth\labelwidth{Fig. 999:}
	\leftmargin\labelwidth
	\advance\leftmargin\labelsep\usecounter{enumi}}}
 \relax
\def\reflist{\section*{References\markboth
	{REFLIST}{REFLIST}}\list
	{[\arabic{enumi}]\hfill}{\settowidth\labelwidth{[999]}
	\leftmargin\labelwidth
	\advance\leftmargin\labelsep\usecounter{enumi}}}
 \relax

\catcode`\@=11

\def\FERMIPUB{}
\def\FERMILABPub#1{\def\FERMIPUB{#1}}
\def\ps@headings{\def\@oddfoot{}\def\@evenfoot{}
\def\@oddhead{\hbox{}\hfill
	\makebox[.5\textwidth]{\raggedright\ignorespaces --\thepage{}--
	\hfill {\rm FERMILAB--Pub--\FERMIPUB}}}
\def\@evenhead{\@oddhead}
\def\subsectionmark##1{\markboth{##1}{}}
}

\ps@headings

\relax
\def\sp#1{{\rm Tr}\biggl( #1 \biggr)}
\def\Ud{U^\dagger}
\def\ra{\rightarrow}

\def\frac#1#2{{#1 \over #2}}
\def\slash#1{\rlap/#1}
\FERMILABPub{93/004--T}
\begin{document}
\begin{titlepage}
\def\ba{\begin{array}}
\def\ea{\end{array}}
\def\thefootnote{\fnsymbol{footnote}}
\begin{flushright}
	FERMILAB--PUB--93/004--T\\
	January 1993
\end{flushright}
\vfill
\begin{center}
{\large \bf RARE  AND RADIATIVE KAON DECAYS}\footnote{To appear in
Annual Review of Nuclear and Particle Science, Volume 43.} \\
\vfill
	{\bf L.~Littenberg$^{(a)}$ and G.~Valencia$^{(b)}$}\\
{\it  $^{(a)}$ Physics Department\\
               Brookhaven National Laboratory\\ Upton, NY 11973}\\
{\it  $^{(b)}$ Theoretical Physics\\
               Fermi National Accelerator Laboratory \\
               Batavia, IL 60510}\\
\vfill
     %	{\large \bf ABSTRACT}
\end{center}
\begin{abstract}

We review the status of theory and experiment of very rare and
forbidden kaon decays.
We then review the radiative non-leptonic decays, and the associated
Dalitz pair modes. We pay particular attention to the study of long distance
physics in radiative decays within
the framework of chiral perturbation theory ($\chi$PT).
We discuss the
experiments that will run in the near future and the modes that they will be
able to study.

\end{abstract}

\end{titlepage}

\tableofcontents

\clearpage

\section{Introduction}
\label{s: intro}

Rare kaon decays offer the possibility of probing
high energy scales by doing precise low energy measurements.
In this case, the precision is obtained by looking at very rare decays.
The advantage of the kaon system is, of course, the long lifetime of both
the $K_L$ and the $K^\pm$. In this paper we review the status of theory and
experiment of rare, forbidden and radiative kaon decays. Other recent reviews
include Refs.~\cite{bryman,rhl,rbcfp,rrw,rdhv,ekre,mar,wolf,buras}.

We concentrate in three distinct classes of decays. We first discuss
a set of modes that does not occur in the
standard model. Studies of these modes constitute searches for new physics,
and in some cases these rare kaon decays are the most sensitive probes
for certain kinds of new interactions.

We then study
those modes that occur in the standard model mostly through short distance
physics, and that are thus amenable to a conventional perturbative
treatment. These modes occur at the one-loop level,
via penguin and box diagrams as discussed in section~\ref{s: sdp}.
The calculation of these
transitions is by now standard\cite{inami,burasb}, and
they are dominated by the top-quark intermediate state.
The contribution of the charm-quark,
as well as perturbative QCD corrections, are also known.
The theoretical aspects of these calculations have been recently
reviewed in the literature by Buras and Harlander, Ref.~\cite{buras}.

The interest of these modes lies in the possibility of measuring
some of the parameters in the standard model. The rates are sensitive
to the values of the top quark mass and of the CKM mixing angles.
For our discussion, we will assume a unitary $3 \times 3$ CKM matrix
as parameterized by Wolfenstein~\cite{wolkm}. The advantage of this
approximate parameterization is that it exhibits the hierarchical
structure of the CKM matrix. It yields expressions that show what is
the mixing angle suppression of a transition. This form of the CKM
matrix is:
\beqn
V =
\left( \begin{array}{ccc}
       1-\lambda^2 / 2  & \lambda   &  A \lambda^3(\rho-i\eta) \\
       -\lambda & 1-\lambda^2 /2   &  A \lambda^2 \\
       A \lambda^3(1-\rho-i\eta) & -A \lambda^2 & 1
       \end{array} \right)
\label{ckmm}
\eeqn
The present knowledge of the mixing angles in the CKM matrix, Eq.~\ref{ckmm},
is summarized in Ref.~\cite{wolf,buras}:
\beqn
\lambda & = &  0.22  \label{knownckm:1}  \\
A & = & 0.9 \pm 0.1 \label{knownckm:2}  \\
\sqrt{\rho^2 + \eta^2} & = & \cases{0.4 \pm 0.2 & \cite{wolf} \cr
                                    0.59 \pm 0.18 & \cite{buras}\cr}
\label{knownckm:3}
\eeqn
Eq.~\ref{knownckm:1} is known from $K_{l3}$ and hyperon decays \cite{vusex},
while Eq.~\ref{knownckm:2} is now usually extracted from analyses of
$B \to D (D^*) e \nu$ decays \cite{vcbex}.  Eq.~\ref{knownckm:3} is
obtained by fitting the observed lepton energy spectrum from B semileptonic
decay as a sum of $b \to c$ and $b \to u$ contributions~\cite{buobc}.
The two different numbers result from the new CLEO value for $V_{ub}$
and the from the old CLEO and ARGUS result.
To obtain separate determinations of $\rho$ and $\eta$, additional input
is necessary.  However, the other currently available sources of
information, such as $\overline B - B$ mixing, $\epsilon$ and $\epsilon'/
\epsilon$ are afflicted both by dependence on the top-quark mass
and calculational difficulties.  Since we expect a direct
measurement of the top-quark mass from collider experiments before
too long, the latter problem is perhaps more severe.  Here
rare kaon decays have the potential to contribute substantially.
The measurement of the decay rates of processes such as
$K^+ \to \pi^+ \nu \overline \nu$ and $K_L \to \pi^0 \nu \overline \nu$ can
provide
constraints on $\rho$ and $\eta$ with very little theoretical uncertainty,
once the top quark mass is measured.  Other measurements
such as that of the parity-violating $\mu^+$ decay asymmetry
in $K^+ \to \pi^+ \mu^+ \mu^-$ can provide complementary information,
given only modest theoretical input.
The first measurements of $\rho$ and $\eta$ will
complete our knowledge of the CKM matrix. Subsequent
determinations of any of the parameters will permit us to
test the three generation structure of the standard model.

Finally we discuss radiative decays that
are expected to be dominated by long distance physics. In this case
we do not have predictions directly from the standard model because
we do not know how to handle the non-perturbative aspects of the
strong interactions. Although we expect that this problem will be
solved at some point by lattice calculations, at present,
the only systematic framework we have to study these modes is
chiral perturbation theory. Within $\chi$PT, we parameterize our
ignorance of strong interaction dynamics in terms of a few unknown coupling
constants. These constants are then measured in some processes,
and after that, they can be used to predict additional decay modes.
We review the basics of this approach in the following section.

\clearpage

\section{Chiral perturbation theory}
\label{s: cpt}

A conventional calculation of non-leptonic kaon decays in the
standard model leads to the evaluation of matrix elements of four-quark
operators between meson states. This is a non-perturbative problem that
remains unsolved. In $\chi$PT one replaces the standard model with
an effective low energy field theory
written directly in terms of meson fields. Effective field
theories contain an infinite number of operators and are, therefore, not
very useful unless one has a way to organize the operators and to
identify the most important ones. $\chi$PT organizes the operators in
the low energy effective Lagrangian in terms of the number of derivatives
(and external fields) that occur. This corresponds to an expansion of
amplitudes in powers of the momentum (or energy) of the external particles.
The energy scale for this expansion is set by the scale of chiral
symmetry breaking, $\Lambda_{CSB}$, empirically about $1~{\rm GeV}$
\cite{cpta,cptb,cptc,dghb}.

The effective Lagrangian is constructed on the basis of chiral symmetry,
an approximate symmetry of the QCD Lagrangian. In the limit of massless
$u$, $d$ and $s$ quarks QCD has a global $SU(3)_L\times SU(3)_R$ chiral
symmetry that is broken spontaneously to $SU(3)_V$. The chiral Lagrangian is
a compact way to keep track of the $SU(3)_V$ relations between amplitudes,
as well as of the relations between amplitudes with different numbers
of pions that follow from the spontaneously broken global symmetry
(the soft pion theorems).

Apart from including all possible operators that are chirally invariant
(organized in terms of number of derivatives), $\chi$PT incorporates
deviations from chiral symmetry due to the small quark masses. This leads
to other operators that are organized as an expansion in powers of meson
masses (and number of derivatives).

Since the typical momentum of particles in a kaon decay is roughly the
kaon mass, we can give a very rough estimate for the size of the corrections
that one can expect in a $\chi$PT calculation: they will be
of order $m^2_K/1~{\rm GeV}^2\sim 25\%$ of the highest order kept in the
calculation.

The framework of $\chi$PT
has proved extremely useful for
analyzing low energy processes involving the pseudoscalar meson octet and
photons. At low energies, the strong and electromagnetic interactions of
these particles can be adequately described with a chiral Lagrangian with
up to four derivatives. The most general chiral Lagrangian to this order has
been written down by Gasser and Leutwyler \cite{cptb}. It consists of two
terms at leading order, ${\cal O} (p^2)$:
\beq
{\cal L}^{(2)}_S = {f^2_\pi \over 4}\sp{D_\mu U D^\mu \Ud}
+ B_0 {f^2_\pi \over 2}\sp{MU+\Ud M}.
\label{slt}
\eeq
$M$ is the diagonal matrix $(m_u,m_d,m_s)$, and
the meson fields are contained in the matrix $U=\exp(2i\phi /f_\pi)$
with:
\beqn
\phi &=& {1 \over \sqrt{2}}
\left( \begin{array}{ccc}
\pi^0 /\sqrt{2}+\eta /\sqrt{6} & \pi^+ & K^+ \\
\pi^- &-\pi^0 /\sqrt{2}+\eta /\sqrt{6} & K^0 \\
K^- & \overline{K^0} & 2 \eta /\sqrt{6}
       \end{array} \right)
\label{pions}
\eeqn
$U$ transforms under the chiral group as $U \ra R U L^\dagger$. We will
restrict ourselves to the case where photons are the only external fields.
In this case the covariant derivative is given by (we will not discuss
radiative semileptonic decays \cite{rsld}):
\beq
D_\mu U = \partial_\mu  U - i e A_\mu [Q,U] .
\label{covd}
\eeq
and $Q$ is the diagonal matrix $(-2/3,1/3,1/3)$. The two constants that
appear at this order are the pion decay constant \cite{holfpi},
\beq
f_\pi =(92.4 \pm 0.2)~{\rm MeV}
\eeq
and the ratio between meson masses and current quark masses $B_0$.
Ignoring isospin breaking, $m_u=m_d=m$, $B_0$ is given by:
\beq
B_0={m^2_{\pi} \over 2m}={m_{K}^2 \over m +m_s}
={3 m^2_\eta \over 2 m + 4m_s}
\label{mudef}
\eeq
At this order, the Gell-Mann-Okubo relation follows:
\beq
3m^2_\eta + m^2_\pi = 4m^2_K
\label{gmor}
\eeq

At next to leading order, ${\cal O}(p^4)$, there are ten more operators
\cite{cptb} without epsilon tensors.
For the processes of interest we will only need two out of these ten terms.
When photons are the only external fields they read:
\beq
{\cal L}^{(4)}_S = -ieL_9 F_{\mu \nu}{\rm Tr}Q\biggl(D^\mu \Ud D^\nu U
+D^\mu U D^\nu \Ud \biggr) +eL_{10}F^{\mu\nu}F_{\mu\nu}\sp{UQ\Ud Q}
\label{slf}
\eeq
and $F_{\mu\nu}$ is the usual electromagnetic field strength tensor.
At this same order there are also terms that contain epsilon tensors. These
have the same origin as the triangle anomaly responsible for $\pi^0 \ra
\gamma \gamma$, and do not involve any unknown coefficients. They are
contained in the Wess-Zumino-Witten anomalous action \cite{wzw},
and the terms of interest to us are:
\beqn
{\cal L}^{(4)}_{WZW} &=& {\alpha \over 8 \pi f_\pi} \epsilon_{\mu\nu\rho\sigma}
F^{\mu\nu}F^{\rho\sigma} \biggl(\pi^0 + {\eta \over \sqrt{3}}
\biggr) \nonumber\\
&-&{i e \over 4\pi^2f_\pi^3} \epsilon_{\mu\nu\rho\sigma} A^\mu
\partial^\nu \pi^+ \partial^\rho \pi^- \partial^\sigma\biggl(\pi^0 +
{\eta \over \sqrt{3}}\biggr)
\label{wzwl}
\eeqn
A complete  calculation to ${\cal O}(p^4)$ consists of tree-level diagrams
with vertices from Eqs.~\ref{slt}, \ref{slf}, and \ref{wzwl}, and
of one-loop diagrams using only Eq.~\ref{slt}.
The divergences that appear in the loop calculation are absorbed by
renormalization of the couplings in Eq.~\ref{slf}.
The renormalized couplings that we will use are defined by
regularizing in $n=4-\epsilon $ dimensions. In terms of
$\lambda_0  \equiv  (2 / \epsilon
+{\rm ln}4\pi +1 - \gamma -{\rm ln}\mu^2 ) /32 \pi^2$, they are
\cite{cptb}:
$L_{9,10}^r(\mu)= L_{9,10} \pm  \lambda_0 / 4$. These, and the other
eight coupling constants have been fixed from experiment \cite{cptb}.
This completes the framework needed to discuss strong and electromagnetic
processes involving the pseudoscalar meson octet and photons.

To study kaon decays we need, in addition, an effective Lagrangian for the weak
interactions. In the standard model, the dominant $|\Delta S| =1$
operators in the effective
weak Hamiltonian transform as $(8_L,1_R)$ or $(27_L,1_R)$ under
chiral rotations. We can write a chiral representation for operators
with these transformation properties, and once again organize
them in terms of the number of derivatives.
The lowest order Lagrangian constructed in this way
contains two derivatives \cite{cron}:
\beqn
{\cal L}^{(2)}_W &=& {G_F \over \sqrt{2}}|V_{ud}V^*_{us}|\biggl[
g_8 \sp{\lambda_6 L_\mu L^\mu} + g_{27}^{|\Delta I| ={1\over 2}}\biggl(
L_{\mu13}L^\mu_{21}+L_{\mu23}(4L^\mu_{11}+5L^\mu_{22})\biggr)
\nonumber \\
&+& g_{27}^{|\Delta I|={3 \over 2}}\biggl(
L_{\mu13}L^\mu_{21}+L_{\mu23}(L^\mu_{11}-L^\mu_{22})
\biggr) \biggr]
\label{wlt}
\eeqn
where $L_\mu =if^2_\pi U D_\mu U^\dagger$. We can use this Lagrangian
to compute $K\ra \pi \pi$ decays, and fit the unknown constants from
experiment \cite{cron,dghpr,devlin}. The result is well known, the
amplitudes with $|\Delta I| = 3/2$ are much smaller than those with
$|\Delta I |= 1/2$. In terms of the couplings of Eq.~\ref{wlt} the
result is:
\beq
{|g_8 +g_{27}^{|\Delta I|={1\over 2}}|\over
|g_{27}^{|\Delta I|={3 \over 2}}|}
\approx {5.1 \over 0.16} \approx 32.
\label{deltai}
\eeq
{}From now on, we will use $|g_8| \approx 5.1$ and drop the terms that
transform as $(27_L,1_R)$. We will also use the notation:
\beq
G_8 \equiv {G_F \over \sqrt{2}}|V_{ud}V^*_{us}|g_8 \approx
9.1 \times 10^{-6}~{\rm GeV}^{-2}.
\label{defge}
\eeq

The situation at next to leading order
is much more complicated: a very large number of
operators, and therefore of unknown coupling constants, has been identified
\cite{kambor}. However, for the radiative decays that we will consider,
only a few of those operators play a role under the following
assumptions \cite{rafao}:
\begin{description}
\item[1.] Octet dominance. We will not include $|\Delta I| = 3/2$
operators.
\item[2.] CPS symmetry. The effective weak Hamiltonian in the standard model
is invariant under a CP transformation followed by the interchange of
$d \leftrightarrow s$. This same symmetry is imposed on the effective
Lagrangian.
\item[3.] Photons are the only external fields, so that
$F^L_{\mu\nu}=F^R_{\mu\nu}=eQF_{\mu\nu}$.
\item[4.] For the normal (odd) intrinsic parity terms (those without
(with) an epsilon
tensor), we will be interested only in terms with at most two
(three) meson fields.
\end{description}
The next to leading order weak Lagrangian then reads \cite{rafao,eprt}:
\beqn
{\cal L}^{(4)}_W &=& -ie {G_8 \over f^2_\pi} F^{\mu\nu}
\biggl[w_1\sp{Q\lambda_6 L_\mu L_\nu} \nonumber \\
&+& w_2 \sp{QL_\mu \lambda_6 L_\nu}
+ ief^4_\pi w_4 F_{\mu\nu}\sp{\lambda_6 QUQ\Ud} \biggr]
\label{wlfn}
\eeqn
for the normal intrinsic parity sector, and \cite{chengan,bepan}:
\beqn
{\cal L}^{(4)}_W &= & ie {G_8 \over f^2_\pi}
\epsilon^{\rho\sigma\mu\nu}{F}_{\rho\sigma}
\biggl[a_1\sp{Q L_\mu}\sp{\lambda_6 L_\nu} \nonumber \\
&+&a_2\sp{U Q\Ud L_\mu}
\sp{\lambda_6 L_\nu}
+ a_3\sp{\lambda_6 [UQ\Ud,L_\mu L_\nu ]} \biggr]
\label{wlfa}
\eeqn
for the odd intrinsic parity sector.

As in the case of the strong interaction Lagrangian, a calculation to
${\cal O}(p^4)$ involves tree and one-loop graphs with Eq.~\ref{wlt}, and
tree graphs with Eqs.~\ref{wlfn}, \ref{wlfa}.
The loop graphs are again divergent, and the divergences are
absorbed by renormalization of the couplings in Eq.~\ref{wlfn} \cite{rafao}:
$w_{1,2}^r(\mu) = w_{1,2} \mp  \lambda_0$,
$w_4^r(\mu) = w_4 - {1 \over 2} \lambda_0$.
After using a few experimental results as input to fit the unknown
coefficients, we can proceed to make predictions for other processes.
A detailed fit of these constants from experiment is in itself interesting,
as it provides a compact parameterization of low energy data that can
be used to compare with first principles calculations when these
become available. In the meantime, they also provide a framework
for systematic tests of different models.

There have been some attempts to derive the effective Lagrangian.
The case of Eq.~\ref{wlt} has been studied in detail using a $1/N_c$
analysis of the strong interactions \cite{bbg}, and a
quark model \cite{goctet}.
There has also been considerable activity using
resonance saturation models \cite{vmdl} and quark models \cite{qml}
to estimate the ${\cal O}(p^4)$ coupling constants.
We summarize our knowledge of the couplings in the strong
interaction sector in Table~\ref{t: scc}
\cite{donho}.

\begin{table}[htb]
\centering
\caption{Values of $L_{9,10}$.}
\begin{tabular}{|l|c|c|} \hline
         &   $L_9^r(\mu=m_\rho)$  &  $L_{10}^r(\mu=m_\rho)$ \\ \hline
Experiment   & $(6.9 \pm 0.2)\times 10^{-3}$
           & $(-5.2\pm 0.3)\times 10^{-3}$  \\ \hline
Vector Dominance & $7.3\times 10^{-3}$
           & $-5.8\times 10^{-3}$  \\ \hline
Quark Models  &  $6.3 \times 10^{-3} $
           & $ -3.2 \times 10^{-3}$\\
\hline
\end{tabular}
\label{t: scc}
\end{table}

The vector dominance model is a tree-level resonance saturation model.
It produces scale independent couplings, and the
resonance saturation assumption is implemented by identifying them
with the running couplings at a scale equal to the resonance mass.
There are many variations of the quark model results, but they all start
from considering a free quark loop. These models are
inspired by the $1/N_c$ expansion, but they are not
complete calculations to leading order in $1/N_c$. They also result
in couplings that are scale independent. We will
identify them with the running couplings at a scale equal to twice
the constituent quark mass, roughly the rho mass (the scale
dependence is non-leading in $1/N_c$ so we are free to choose any scale).
Attempts to incorporate gluonic
corrections seem to improve the agreement with experiment, although
precise quantitative predictions are not available \cite{qml}.

For the weak interaction parameters there are also several models.
Among them the weak deformation model \cite{wdmw}, and quark models
\cite{hyc,bruno}.
We summarize these results in Table~\ref{t: wcc}.

\begin{table}[htb]
\centering
\caption{Model calculations of $w_i$}
\begin{tabular}{|l|c|c|c|} \hline
&  $w_1$  &  $w_2$  & $w_4$  \\ \hline
weak deformation model \cite{wdmw}
          & $0.007$ & $0.028$ & $-0.021$  \\ \hline
factorization \cite{hyc}
         & $0.025$ & $0.025$ & 0.0  \\ \hline
quark model \cite{bruno}
         & $-0.003$ & $0.013$ & $-0.005$   \\ \hline
\end{tabular}
\label{t: wcc}
\end{table}

\noindent Similarly, the large-$N_c$ factorizable contributions to
the constants in Eq.~\ref{wlfa} are \cite{chengan,bepan}:
\beq
a_1={1 \over 4 \pi^2},\;a_2={1 \over 8 \pi^2},\;a_3={1 \over 16 \pi^2}.
\eeq

For some of the modes that we will discuss, we will need
the matrix elements of currents; in analogy with the
semileptonic decays. For these we will use the following form of the
currents in terms of mesons:
\beqn
\overline{s} \gamma_\mu u & \ra &
 -i \biggl[ \biggl(
\pi^+\partial_\mu K^0-K^0\partial_\mu \pi^+\biggr)+{1\over \sqrt{2}}\biggl(
\pi^0\partial_\mu K^+ - K^+ \partial_\mu \pi^0 \biggr) \biggr]
\nonumber \\
\overline{s} \gamma_\mu \gamma_5 u &\ra &
- \sqrt{2}f_\pi \partial_\mu K^+
\label{currcl}
\eeqn
for the charged current, and
\beqn
\overline{s}\gamma_\mu d & \ra & -i\biggl[\biggl(\pi^-\partial_\mu K^+
- K^+\partial_\mu \pi^-\biggr) + {1 \over \sqrt{2}}\biggl(
K^0 \partial_\mu \pi^0 - \pi^0\partial_\mu K^0\biggr)\biggr] \nonumber \\
\overline{s} \gamma_\mu \gamma_5 d & \ra & -\sqrt{2}f_\pi \partial_\mu K^0
\label{curbos}
\eeqn
for the neutral current. These results follow from Eq.~\ref{slt}~\cite{cptc},
and are valid to lowest order in $\chi$PT. We have only kept
terms with up to two mesons. From Eq.~\ref{curbos}
we can also find expressions for the scalar and pseudoscalar
densities:
\beqn
\overline{s}d & \ra & i B_0 \biggl(\pi^- K^+ + {1 \over \sqrt{2}}K^0
\pi^0\biggr)
\nonumber \\
\overline{s} \gamma_5 d & \ra & \sqrt{2} f_\pi B_0 K^0
\label{denbos}
\eeqn
$B_0$ is the same as that in Eq.~\ref{mudef}.
Although $B_0$ is not a physical quantity, we will simply use
$B_0 / m_\mu \approx 15$.
We have divided by the muon mass for convenience.
This number reflects
that matrix elements of scalar operators
are somewhat enhanced.

\clearpage

\section{Lepton family number violating decays}
\label{s: lfnv}

In the minimal standard model with massless neutrinos, the lepton
family number is absolutely conserved so these decays do not occur.
The observation of $K_L \ra \mu^\pm e^\mp$ or $K \ra \pi \mu e$ decays
would therefore be a clear indication of new physics.

A model independent study of this type of processes can be done following the
approach of Buchm\"{u}ller and Wyler, Ref.~\cite{buwy}.
The physics beyond the standard
model is parameterized by an effective Lagrangian that is gauge invariant
under $SU(3)_c\times SU(2)_L\times U(1)_Y$. The leading effective Lagrangian,
relevant for our discussion, is given by the sum of four
fermion operators of the form:
\beqn
{\cal O}_{V-A}&=& C_{V-A} {g_n^2 \over  \Lambda^2}\overline{\mu}
\gamma_\mu {(1+\gamma_5 )\over 2} e \overline{s}
\gamma^\mu {(1+\gamma_5 )\over 2} d \nonumber \\
{\cal O}_{S\pm P}&=& C_{S\pm P} {g_n^2 \over  \Lambda^2}\overline{\mu}
{(1-\gamma_5 )\over 2} e \overline{s}
{(1+\gamma_5 )\over 2} d
\label{efop}
\eeqn
We have included a (weak) gauge coupling $g_n^2$ to reflect the fact that
we think of these operators as originating in the exchange of a heavy
gauge boson (or perhaps a scalar) in the new physics sector. We will take
$g_n =g_2$ for simplicity and absorb any difference, as well as any
mixing angles or other factors into the coefficient $C_i$.
The factors of 2 have been included for later convenience. In the absence of
any dynamical information to the contrary, it is natural to assume that
$C_i$ is of order ${\cal O}(1)$. $\Lambda$ is the scale that characterizes the
heavy degrees of freedom, typically the mass of the exchanged boson. It
is then conventional to take $C_i=1$ and interpret the bounds on the decays
induced by these operators as bounds on the ``scale of new physics''
$\Lambda$. We have not listed all possible Dirac structures in Eq.~\ref{efop},
but just two illustrative ones. The operators could be purely vector or
axial-vector operators, for example. Tensor operators, however, can be
reduced to the others by Fierz rearrangements~\cite{buwy}.

We compare the rates induced by these operators with Standard Model rates
using the S.M. effective four-fermion operator for semileptonic decays:
\beq
{\cal L}= {g^2_2 \over {2 m^2_W}}V_{us} \overline{\nu}\gamma_\mu
{(1+\gamma_5)\over 2}\mu \overline{s} \gamma_\mu {(1+\gamma_5 )\over 2}u
\eeq
and using Eqs.~\ref{currcl},~\ref{curbos},~\ref{denbos}.

Models that violate lepton flavor number will also induce processes
like $\mu \ra e \gamma$, $\mu^\pm \ra e^\pm e^+ e^-$,
$\mu \ra e$ conversion in the field of a heavy nucleus,
and $\Delta M(K_L - K_S)$ among others. For the latter
one finds, $\Delta M = f^2_K m_K / 3 \Lambda^2$ \cite{buwy}.
The experimental
value $\Delta M(K_L-K_S)=(3.522\pm 0.016)\times 10^{-12}~{\rm MeV}$ \cite{pdb},
places a bound $\Lambda > 830~{\rm TeV}$ for $(V-A)\otimes (V-A)$ operators.
This bound will be better than the one obtained from $K_L \ra \mu^\pm e^\mp$
until this process reaches a sensitivity around $10^{-15}$. However,
this comparison of different processes assumes that all the coefficients
$C_i$ in the effective Lagrangian are of the same order.
Although this is a natural assumption, it may not be true for given
coefficients in specific models. Therefore,
it is important to study all the different processes, since at some level
they provide complementary information.

\subsection{$K_L \ra \mu^\pm e^\mp$}

It is standard to compare this mode to the rate for $K^+ \ra \mu^+ \nu$,
since in the limit of vanishing electron mass it has the same kinematics.
Given the pseudoscalar nature of the kaon, only the axial vector quark
current or pseudoscalar density can contribute, as can
be seen from Eqs.~\ref{curbos}, \ref{denbos}. One finds:
\beq
{\Gamma (K_L \ra \mu^+ e^-) \over \Gamma(K^+ \ra \mu^+ \nu)} =
2{C^2_{V-A}\over |V_{us}|^2}
\biggl({g_n \over g_2}\biggr)^4\biggl({m_W \over \Lambda}\biggr)^4
\label{rmeva}
\eeq
for $V-A$ operators, and
\beq
{\Gamma (K_L \ra \mu^+ e^-) \over \Gamma(K^+ \ra \mu^+ \nu)} =
2{C^2_{S\pm P}\over |V_{us}|^2}
\biggl({g_n \over g_2}\biggr)^4\biggl({m_W \over \Lambda} \biggr)^4
\biggl( {B_0 \over m_\mu}\biggr)^2
\label{rmeps}
\eeq
for scalar, pseudoscalar operators.

Fig.~\ref{f: lfv} a and b show the recent results in the search for
$K_L \ra \mu e$.  The lack of a signal in the region corresponding
to $M(\mu e) \sim M_K$ with aligned initial and final states allows
$90 \%~c.l.$ upper limits to be set.  These are $B(K_L \ra \mu^\pm e^\mp)
< 9.4 \times 10^{-11}$ for KEK-137 \cite{mue137} and
$B(K_L \ra \mu^\pm e^\mp) < 3.3 \times 10^{-11}$ for AGS-791 \cite{mue791}.
For purely left handed operators, the latter experimental limit
places the bound $\Lambda > 108~{\rm TeV}$. For scalar
operators as in Eq.~\ref{efop}, we find from Eq.~\ref{rmeps}
$\Lambda > 420~{\rm TeV}$. These results can be interpreted as bounds on the
mass of new particles in different models. For example, we identify
$\Lambda \ra \sqrt{2}m_H$; the mass of a $\Delta G =0$ horizontal gauge
boson as discussed in Ref.~\cite{cahara}, to obtain $m_H > 77~{\rm TeV}$.
The bound on the family replication model of Ref.~\cite{pati} is placed
by the scalar operators. Using Eq.~\ref{rmeps} and $g_n^2/\Lambda^2 \ra
4 \sqrt{2}g^2_{hc}\sin^4\alpha / \Lambda^2_0$ we find $\Lambda_0 > 10^5 ~{\rm
TeV}$
when we use the parameters $\sin\alpha = 0.04$ and $g_{hc} =1$. Similarly,
we can use Eq.~\ref{rmeps} to compare with the scalar leptoquark operators
of Ref.~\cite{shanker} to obtain $m_H > 15~{\rm TeV}$, and $m_P > 2.8 ~{\rm
TeV}$ for
pseudoscalar leptoquarks. In this case the bounds on the particle mass
are not as strong as implied by Eq.~\ref{rmeps} because the couplings
$C_{S\pm P}$ in
these models are suppressed by small mass ratios $m_f / m_W$.

One can also use the form of the new operators Eq.~\ref{efop} to compare
the reach of the rare kaon decay experiments with that of rare B decays.
For example, taking a purely $(V-A)\otimes(V-A)$ operator one finds
\cite{mar}:
\beq
{B(B^0 \ra \mu^\pm e^\mp) \over B(K_L \ra \mu^\pm e^\mp)} =
\biggl({C^B_{V-A} \over C^K_{V-A}}\biggr)^2{m_B \over m_K}
{\tau_B \over \tau_{K_L}} \approx 3 \times 10^{-4}
\eeq
In the last step we have assumed that the
$B$ and $K$ coupling constants are of the same
order. This indicates that rare kaon decays are more
likely to be sensitive to this kind of new physics than rare B decays.
Of course, the situation may be different for other operators.

These decays are also allowed in minimal extensions of the Standard Model
in which the neutrinos are given a mass. The decays then proceed via
box diagrams as in Fig.~\ref{f: bnn}. The decay rate for $K_L \ra \mu^\pm
e^\mp$ is proportional to the product of mixing angles between the $\mu$, $e$
and the heavy neutral lepton $N$: $|U_{Ne}U^*_{N\mu}|^2$. Using the
result $B(\mu \ra e \gamma)< 4.9 \times 10^{-11}$ \cite{mueexp}, the authors
of Ref.~\cite{acpak} find $|U_{Ne}U^*_{N\mu}|^2 < 7 \times 10^{-6}$ for
$m_N > 45~{\rm GeV}$. From this they conclude that
$B(K_L \ra \mu^\pm e^\mp)$ is at most
$10^{-15}$ in this type of models. Marciano \cite{marhn} has pointed
out that there is a better bound  $|U_{Ne}U^*_{N\mu}|^2 < 10^{-8}$
coming from $ \mu  \ra e $ conversion in the field of a heavy nucleus.
With this bound one finds $B(K_L \ra \mu^\pm e^\mp)$ to be at most
a few times $10^{-18}$. By using the additional theoretical prejudice
that the mixing angles should be proportional to mass ratios
of the form $m_e / m_N$, the authors of
Ref.~\cite{babrbe,lasasc} find the much stronger
limit $B(K_L \ra \mu^\pm e^\mp)< 10^{-25}$. At this level the decay
is completely unobservable.
If, in addition, one considers left-right symmetric models,
the authors of Ref.~\cite{babrbe,lasasc} find that the rate
$B(K_L \ra \mu^\pm e^\mp)$ can be as
large as $10^{-13}$, although this happens only
in a small corner of parameter space.

There is an ultimate background to the decays $K_L \ra \mu^\pm e^\mp$. It
is given by the standard model process $K_L \ra \mu^\pm e^\mp \nu_\mu \nu_e$.
To leading order in $\chi$PT this process is dominated by
$K_L \ra \pi^\pm e^\mp \nu_e$ followed by $\pi^\pm \ra \mu^\pm \nu_\mu$. The
branching ratio for this chain can be estimated using the narrow
width approximation to be a huge $38\%$. However, it is easy to see
that the maximum invariant mass of the lepton pair is $m_{\mu e}
< 489~{\rm MeV}$. It is therefore possible to remove this background with a cut
on the lepton pair invariant mass. Going beyond the narrow width
approximation, and including next to leading order terms in $\chi$PT
can yield a lepton pair invariant mass larger than $489~{\rm MeV}$. However,
after the $489~{\rm MeV}$ cut is imposed, the rate is reduced to the
$10^{-23}$ level \cite{dava}.

\subsection{$K^+ \ra \pi^+ \mu^\pm e^\mp$}

In this case we compare the rate to that of $K^+ \ra \pi^0 \mu^+ \nu$.
{}From Eqs.~\ref{curbos}, \ref{denbos} we can see that this mode is
only sensitive to the vector quark current or scalar density. In general,
this mode is thus probing different operators than the previous one.
For $V-A$ operators we find:
\beq
{\Gamma (K^+ \ra \pi^+ \mu^+ e^-) \over \Gamma(K^+ \ra \pi^0 \mu^+ \nu)} =
8{C^2_{V-A}\over |V_{us}|^2}
\biggl({g_n \over g_2}\biggr)^4\biggl({m_W \over \Lambda}\biggr)^4
\eeq
Fig.~\ref{f: lfv} c shows the data from the most recent search for
$K^+ \ra \pi^+ \mu^+ e^-$ \cite{alee}.  The lack of candidates within
the search region allows the $90 \%~c.l.$ upper bound
$B(K^+ \ra \pi^+ \mu^+ e^-)< 2.1 \times 10^{-10}$ to be set.  The
corresponding bound on the state with opposite lepton charges,
$B(K^+ \ra \pi^+ \mu^- e^+)< 6.9 \times 10^{-9}$ \cite{diam}, was
set in an earlier experiment.
Using these experimental bounds we find $\Lambda > 76~{\rm TeV}$.
Once again, we can use this result to place bounds on masses of new
particles within  specific models.

Given the large number of parameters in extensions of the standard model,
it is very difficult to make definite predictions for any of the rare
decays. Most of these models, however, will give rise to both
$K \ra \mu e$ and $K \ra \pi \mu e$, as well as additional
contributions to processes that do occur in the minimal standard model
such as $K \ra \pi \nu \overline{\nu}$, $K_L \ra \ell^+ \ell^-$ and $\Delta M
(K_L - K_S)$. Discussions of how to use the experimental limits on
all these processes to constrain the parameters in the models can be
found in Ref.~\cite{beyondsm}.

\clearpage

\section{Short distance dominated processes}
\label{s: sdp}

These are modes that can be computed reliably from the standard model,
because they are
not affected significantly by the non-perturbative aspects of the
strong interactions. Within the standard model they are sensitive
to the CKM parameters $\rho$ and $\eta$, and to the top quark mass.
Physics beyond the standard model, as in the models that gave rise
to the modes studied in the previous section, can also affect
these decay modes. We will concentrate on the standard model, but,
as we will see, current experiments have not yet reached the
sensitivity required to observe the standard model rates. This
means that for the time being, these modes constitute a window to
possible new physics.

\subsection{$K^+\rightarrow \pi^+ \nu \overline{\nu}$}

In the standard model this process is mediated by the electroweak penguin
and box diagrams depicted schematically in Fig.~\ref{f: sdd}.
The top-quark contribution is calculated to be \cite{inami}:
\beq
M=A^2\lambda^5(1-\rho-i\eta)
{G_F \over \sqrt{2}}{\alpha \over 2\pi \sin^2\theta_W}
X(x_t) <\pi^+|\overline{s}\gamma_\mu d|K^+>\overline{\nu}\gamma^\mu(1+\gamma_5
)\nu
\label{kptop}
\eeq
where $x_t=m_t^2/m_W^2$ and:
\beq
X(x_t)={x_t \over 8}\biggl({x_t+2 \over x_t-1}+
{3x_t-6\over (x_t-1)^2}{\rm ln}x_t\biggr)
\label{ffxt}
\eeq
Ref.~\cite{burasb} gives the approximate result
$X(x_t)\approx 0.650 x_t^{0.59}$.
The hadronic matrix element can be related by isospin to that occurring in
$K^+ \ra \pi^0 e^+ \nu$ (equivalently using Eq.~\ref{curbos}),
and this allows one to write:
\beq
{B(K^+\ra \pi^+ \nu \overline{\nu}) \over B(K^+ \ra \pi^0 e^+ \nu)}
= \biggl({\alpha \over \pi\sin^2\theta_W}\biggr)^2
A^4 \lambda^8 \biggl[ (1-\rho)^2 + \eta^2\biggr] \biggl[ X(x_t)
\biggr]^2
\label{kpbrt}
\eeq
for each neutrino flavor. The contribution of the charm quark intermediate
state has also been computed. In this case, however, additional
kinematical factors for the lepton masses, as well as QCD corrections
do not allow us to write a simple expression.
The authors of Ref.~\cite{burasb} provide us with an approximate
expression for the total rate with three generations of neutrinos:
\beq
B(K^+\rightarrow \pi^+ \nu \overline{\nu})
\approx 1.97\times 10^{-11} A^4 x_t^{1.18}
[\eta^2+(\overline{\rho_0}-\rho)^2]
\label{kpapp}
\eeq
The parameter $\overline{\rho_0}$ measures the charm quark
contribution and is given in Ref.~\cite{burasb}, a typical
value for it being 1.5. This same reference finds $B(K^+ \ra \pi^+ \nu
\overline{\nu})$ to be in the range $(0.5 - 8.0)\times 10^{-10}$ when all
uncertainties are included.

One can think of long distance contributions to this process that occur
via $\mu$ pole diagrams as in Fig.~\ref{f: impd},
but these have been estimated to be at the
level of $10^{-15}$, much smaller than the short distance
contributions \cite{rhl}.

It has been argued that minimal extensions of the standard model are
unlikely to affect this process significantly, so that its main
interest remains the constraining the CKM parameters \cite{bigi}.
However, even in Ref.\cite{bigi}, several less minimal extensions
were discussed in which the current experimental upper bound
could easily be saturated.  In addition experiments sensitive to
$K^+ \ra \pi^+ \nu \overline{\nu}$ can also see $K^+ \ra \pi^+ X^0$
where $X^0$ is a new light weakly-interacting particle.  Candidates
for such an $X^0$ include axions~\cite{axion}, familons~\cite{familon},
and hyperphotons~\cite{hyper}.  Thus the possibility of novelty in the
topology $K^+ \ra \pi^+ +$~``nothing'' should not be forgotten.

Fig.~\ref{f: 787dat}, shows recent data from AGS-787.
There were separate analyses for the kinematic regions
with $\pi^+$ momenta above and below that corresponding to the
background $K^+ \ra \pi^+ \pi^0$ process.  No candidates were
found within the search regions.  The $90\%$ c.l. upper limits
extracted were $B(K^+ \ra \pi^+ \nu \overline{\nu}) < 7.5 \times 10^{-9}$
and $ 1.7 \times 10^{-8}$ respectively\cite{pnn787,pnn2}.
Assuming a vector momentum spectrum for the $\pi^+$, these can be
combined into an overall upper limit  $B(K^+ \ra \pi^+ \nu \overline{\nu})
< 5.2 \times 10^{-9}$.  The corresponding limit on $K^+ \ra \pi^+ X^0$
for a massless, long-lived $X^0$ is  $1.7 \times 10^{-9}$.   An upgrade
of this experiment expects to reach a sensitivity of $10^{-10}$, at which
point one can begin to extract useful constraints on $\rho$ and $\eta$.

\subsection{$K_L \rightarrow \pi^0 \nu \overline{\nu}$}

The neutral version of the previous mode is completely dominated by
the top-quark intermediate state. The result
for three neutrino species can be written as \cite{burasb}:
\beq
{B(K_L \rightarrow \pi^0 \nu \overline{\nu})  \over
B(K^+ \ra \pi^0 e^+ \nu) }
= {3 \over 2} {\tau_{K_L}
\over \tau_{K^+}}  \biggl( {\alpha \over
\pi \sin^2\theta_W }\biggr)^2 A^4 \lambda^8 \eta^2 \biggl[X(x_t)\biggr]^2
\label{kpapn}
\eeq
The expected rate in the standard model is of order
$10^{-10}$, and this decay would directly measure $\eta$ \cite{litt}. The
potential long distance contributions to this decay are expected
to be very small. Recently, the QCD corrections to the top-quark
contribution to this decay (and to the charged mode as well) have been
computed in Ref.~\cite{bunew}. This calculation reduces the
uncertainty in the rates due to the dependence of the top mass on the
renormalization scale from ${\cal O}(25\%)$ to ${\cal O}(3\%)$. This
makes this mode a particularly clean one to measure $\eta$ within the
standard model.  In principle this measurement can also be combined
with that of the charged mode to reduce uncertainties due to $m_t$
and to $A$.

At present the best limit comes from FNAL-731 \cite{pnn731}, it is
$B(K_L \ra \pi^0 \nu \overline{\nu})<2.2 \times 10^{-4}$.  In this
experiment the decay was sought in the Dalitz mode where the $\pi^0$
decays to $e^+ e^- \gamma$.  Using this technique, FNAL-799
expects to reach sensitivity in the $10^{-8}$ region.  The KAMI
conceptual design report \cite{kami} claims an eventual sensitivity
of a few $\times 10^{-12}$ for this decay at the FNAL Main Injector.
There is also a letter of intent for an experiment \cite{tini} at KEK
with an eventual sensitivity of $<10^{-11}$/event.   Other versions of
this experiment are under consideration at a number of institutions.

\clearpage

\section{Long distance dominated radiative decays}
\label{s: krd}

In this section we consider radiative decays and some of the Dalitz
pair modes associated with them. These decays are sensitive to the
non-perturbative aspects of the strong interactions, and thus they cannot be
predicted reliably from the standard model at present. We resort to the
framework of $\chi$PT to study these modes. As we will see, the
current state of the art, ${\cal O}(p^4)$ calculations, is not
always sufficient for an adequate description of these modes. These modes
are interesting in their own right because they yield information on
the long distance strong interactions. In some cases, a detailed
understanding of these modes is also necessary in the analyses of
other modes that look for CP violation, CKM angles, or new physics.

\subsection{$K_S \ra \gamma \gamma$}

The processes $K_S \ra \gamma \gamma$ and $K_L \ra \pi^0 \gamma \gamma$
share the remarkable feature of being independent of
the couplings in Eqs.~\ref{wlfn}, \ref{wlfa} at ${\cal O}(p^4)$. They are
therefore among the cleanest predictions of $\chi$PT. They can be studied
by considering the diagrams shown in Fig.~\ref{f: kgg}.
A considerable simplification in the calculation is obtained when one
uses the ``diagonal'' basis of Ecker {\it et. al.}, Ref.~\cite{epro}.
Defining $z=q_2^2 /m^2_K$, $r_\pi^2=m_\pi^2 /m_K^2$, we write
the amplitude for $K_1^0 \ra \gamma(q_1) \gamma^*(q_2)$ as restricted by gauge
invariance:
\beq
M_\nu= \epsilon^\mu(q_1)\biggl({m_K^2 \over 2}(z-1)
g_{\mu\nu}+q_{2\mu}q_{1\nu}\biggr)
b(0,z).
\label{giks}
\eeq
The ${\cal O}(p^4)$ result is given by \cite{eprt}:
\beq
b(0,z)={\alpha \over \pi} G_8 2\sqrt{2}f_\pi(1-r^2_\pi)H(z).
\label{ksff}
\eeq
The function $H(z)$ can be written as:
\beq
H(z) = {1 \over 2(1-z)^2}\biggl\{zF\biggl({z\over r_\pi^2}\biggr)
-F\biggl({1\over r^2_\pi}\biggr)-2z\biggl[
G\biggl({z\over r_\pi^2}\biggr)-G\biggl({1\over r^2_\pi}\biggr)\biggr]\biggr\};
\label{hdef}
\eeq
the function $F(x)$ is given by:
\beq
F(x)=\cases{
1-\frac{4}{x}\biggl({\rm arcsin}{\sqrt{x}\over 2}\biggr)^2  &$x\le 4$\cr
1+{1\over x}\biggl(\log  {1- \sqrt{1-4/x}\over 1+\sqrt{1-4/x}}
  +i \pi\biggr)^2 & $x > 4$ ;\cr }
\label{fdef}
\eeq
and $G(x)$ is given by:
\beq
G(x)=\cases{
\sqrt{{4 \over x}-1}\biggl({\rm arcsin}{\sqrt{x}\over 2}\biggr)^2  &$x\le 4$\cr
{1 \over 2}\sqrt{1-{4 \over x}}
\biggl(\log  {1+ \sqrt{1-4/x}\over 1-\sqrt{1-4/x}}
-i \pi\biggr)^2 & $x > 4$ .\cr }
\label{gdef}
\eeq

The situation in $K_S \rightarrow \gamma \gamma$ looks very good,
with $z=0$ in Eq.~\ref{ksff}, one finds
a branching ratio of $2.0 \times 10^{-6}$ \cite{goity}
and the NA31 measurement is
$(2.4\pm 1.2)\times 10^{-6}$ \cite{kssna}.
However, we note the very large
error in the data. Given the fact that the theoretical prediction is rather
clean, it is very important to reduce the experimental error. It is also
possible to study this prediction in the related
Dalitz pair decays $K_1^0 \ra \ell^+ \ell^- \gamma$, for which
one finds \cite{eprt}:
\beq
{d\Gamma \over dz}={\alpha \over 3 \pi}
\Gamma(K_1^0 \ra \gamma \gamma) {2\over z}
(1-z)^3\biggl|{H(z) \over H(0)}\biggr|^2(1+2{r^2_\ell}/z)
\sqrt{1-4r^2_\ell/z}
\label{ksglldis}
\eeq
Ignoring CP violation the predicted branching ratios are then:
\beq
{\Gamma(K_S \ra \ell^+ \ell^- \gamma)\over \Gamma(K_S \ra \gamma
\gamma)}=\cases{
0.016  & $\ell=e$ \cr
3.75 \times 10^{-4} & $\ell=\mu$ \cr}
\label{ksllgres}
\eeq

\subsection{$K_L \ra \pi^0 \gamma \gamma$}

The amplitude for $K_L(p) \rightarrow \pi^0(p^\prime) \gamma
(q_1,\epsilon_1) \gamma (q_2,\epsilon_2)$;
$M= \epsilon_1^{\mu}(q_1) \epsilon_2^{\nu}(q_2) M_{\mu \nu}$,
is restricted by gauge invariance and CP conservation to be of the form:
\beqn
M_{\mu \nu}&=& A(z,\nu)\biggl( -m_K^2 {z \over 2}g_{\mu \nu}+q_{2\mu} q_{1\nu}
\biggr)+ \nonumber\\
&& B(z,\nu)\biggl(-m_K^2 x_1 x_2 g_{\mu \nu}
-{z \over 2} p_{\mu} p_{\nu} + x_1 q_{2 \mu}p_{\nu} +x_2 p_{\mu}q_{1
\nu}\biggr)
\label{kpgg}
\eeqn
where $x_i=p\cdot q_i / m_K^2$,
$z=2q_1 \cdot q_2 /m_K^2$, and $\nu = x_1 -x_2$.
The lowest order amplitude $K_L \rightarrow \pi^0 \gamma \gamma$,
${\cal O}(p^4)$, obtained
from chiral loops gives rise only to the form factor $A(z,\nu)$\cite{epro}.
To this lowest order result we can add the next order terms, ${\cal O}(E^6)$,
as they appear in some models. They can be parameterized in terms of a
single constant $a_V$, and the combined result is:
\beqn
A(z,\nu)&=&
\frac{\alpha}{\pi} G_8 \biggl[ F\biggl(\frac{z}
{r^2_{\pi}}\biggr) \biggl(1-\frac{r^2_{\pi}}{z}\biggr)
+ F(z)\biggl(\frac{1+r^2_\pi}{z}-1\biggr)
+a_V\biggl(3+r_\pi^2-z\biggr) \biggr] \nonumber\\
B(z,\nu)&=&-4 a_V \frac{\alpha}{\pi} G_8 \quad .
\label{pocl}
\eeqn
The function $F(x)$ is given in Eq.~\ref{fdef}.

The constant $a_V$ has been calculated in several models. The simplest
ones are those that consider only pole diagrams for the $E^6$ terms,
such as those of Fig.~\ref{f: poles}.
One should
also consider direct weak counterterms as depicted schematically in
Fig.~\ref{f: poles}. A model to compute these
direct counterterms is the ``weak deformation model'' of \cite{wdmw}.
For this mode, the model predicts the direct weak counterterm contribution
to $a_V$ to be twice as large as that from
the pole terms and to have the opposite sign. The net effect is thus to
change the sign of the constant $a_V$ calculated from pole diagrams alone.
\beq
a_V=\cases{0.32 & VMD (poles); $-0.32$ (poles $+$ WDM) \cite{wdmw} \cr
0.22 & Q.M. (poles); $-0.22$ (poles $+$ WDM) \cite{bdv} \cr}
\label{acqm}
\eeq

We compare these results with experiment \cite{pggNA31,pgg731}
in Fig.~\ref{f: rkpgg}.  In Fig.~\ref{f: rkpgg} a, the various
predictions have been mutually normalized.  It is apparent in
Fig.~\ref{f: rkpgg} b, that the shape of the data of
Ref.~\cite{pggNA31} is well fit by the ${\cal O}(p^4)$ prediction
(this is also true of the data of Ref.~\cite{pgg731}).
The shape of the distribution is affected very little by the small $p^6$
corrections. The corrections (both in the quark model and in the vector
dominance model) tend to increase the number of events to be expected
for the lower values of the photon pair invariant mass. However, this
is a very small effect, and it is difficult to check given the currently
available statistics.

The branching fraction with a cut on the invariant mass of
the photon pair, $M_{\gamma
\gamma}>280~{\rm MeV}$, is \cite{pggNA31,pgg731}:
\beq
B(K_L\rightarrow \pi^0\gamma
\gamma)=\cases{
(0.57^{+0.11}_{-0.06})\times 10^{-6} & $\chi$PT,\cr
(1.7\pm 0.3) \times 10^{-6} & NA31,\cr
(1.86 \pm 0.6) \times 10^{-6} & FNAL-731.\cr}
\label{brscut}
\eeq
The central value for the theory number corresponds to the ${\cal O}(p^4)$
result. The error represents the variation obtained by including the
${\cal O}(p^6)$ terms with $a_V$ from Eq.~\ref{acqm}.
The predictions are significantly smaller than the data. Other models
(outside the realm of $\chi$PT), have predicted larger branching ratios
\cite{segg}; however, they predict a $\gamma \gamma$ spectrum that
strongly disagrees with experiment, whereas $\chi$PT appears
to reproduce the data very well (see Fig.~\ref{f: rkpgg}). These other models
include vector meson exchange diagrams such as those of Fig.~\ref{f: poles}.
This has created the incorrect impression in the literature that it is the
inclusion of vector mesons that is responsible for the large rate.
That this is not so can be seen, for example, in the
model of Ref.~\cite{wdmw}, Eq.~\ref{acqm}, that includes the vector
mesons as well. As noted in Ref.~\cite{bdv}, what leads to large rates
in the models of Ref.~\cite{segg} is the very specific form in which
they include the $\eta^\prime$ pole. This pole, unlike the $\pi^0$ and
$\eta$ poles of Fig.~\ref{f: poles}, cannot be treated unambiguously at
present. It enters at the same order as $\eta -\eta^\prime$ mixing and
other higher order $SU(3)_V$ breaking effects. The authors of
Ref.~\cite{segg} include only some of these effects, and the rate is
very sensitive to the precise way in which this is done. The very large
model dependence introduced by these terms can be appreciated in the mode
$K_L \ra \gamma \gamma$ where a similar situation occurs \cite{lin}.

{}From the shape of the spectrum, NA31 has derived the bound \cite{pggNA31}:
\beq
-0.32 < a_V < 0.19
\label{avbou}
\eeq
at 90\%
confidence level. One must notice, however, that with this range for $a_V$,
the branching ratio (with $M_{\gamma\gamma} > 280~{\rm MeV}$) is not
reproduced:
\beq
0.53 \times 10^{-6} \leq B(K_L \ra \pi^0 \gamma \gamma) \leq
0.68 \times 10^{-6}.
\eeq
Conversely, a fit to the branching ratio yields $a_V \approx -2.0$, which
does not reproduce the shape of the spectrum.
The shape of the spectrum  appears to indicate
that the $\chi$PT arguments are correct, and that
models with very large D-wave amplitudes are
ruled out. The discrepancy in the branching ratio tells us that this
mode will require more theoretical and experimental efforts
in the years to come.

The contribution of the $\Delta I = 3/2$ amplitudes has been recently
estimated in Ref.~\cite{cami}. These authors found that it introduced a small
${\cal O}(10\%)$ correction that tends to make the branching ratio smaller.
The amplitude computed in Ref.~\cite{cami} contributes only to the $A(z,\nu)$
form factor, so the spectrum still has the shape given
by lowest order $\chi$PT.
These same authors have studied possible effects of higher order terms
by using the experimental amplitude for the $K \ra 3 \pi$ vertex. They
find an enhancement of about 26\% in the branching ratio, that
comes mostly from contributions to $A(z,\nu)$.
Although this does go in the right direction, it is too small to account
for the discrepancy with experiment.
We should also remember that the naive estimate for the size of $p^6$
corrections to a $p^4$ amplitude is about $25\%$ as well.

\subsection{$K^+ \ra \pi^+ \gamma \gamma$}

This process is similar to the neutral mode in that the one-loop amplitude
in $\chi$PT is finite. Unlike the neutral mode, however, this mode does
receive contributions at order ${\cal O}(E^4)$ from local counterterms.
An additional feature is the possibility of looking for CP violation
by comparing the two charged modes.
The amplitude for this process is restricted by gauge
invariance to be of the form:
\beq
M=\epsilon^\mu_1\epsilon^\nu_2\biggl[A(z,\nu)\biggl(
-m^2_K{z \over 2}g_{\mu\nu}+q_{2\mu}q_{1\nu}\biggr)
+C(z,\nu) \epsilon_{\mu\nu\rho\sigma}q^\rho_1q^\sigma_2 \biggr]
\label{gicha}
\eeq
The form factor $A(z,\nu)$ to ${\cal O}(p^4)$ is given by \cite{eprt},
\beqn
A(z,\nu) &=& {\alpha \over \pi} G_8 {1 \over 2z} \biggl[
\biggl(r^2_\pi -1-z\biggr)F\biggl({z \over r^2_\pi}\biggr)
+ \biggl(1-z-r^2_\pi\biggr)F(z) + \hat{c} z \biggr] \nonumber \\
\hat{c} &=& 32\pi^2\biggl(4(L_9+L_{10})-{1 \over 3}(w_1+2w_2+2w_4)\biggr)
\label{kpppgg}
\eeqn
The constant $\hat{c}$ is a scale independent combination of couplings in the
effective Lagrangian, reflecting the fact that the loop contributions to this
process are finite. Its real part, ${\rm Re}\hat{c}$, has been
estimated in several models to be:
$0.0$ \cite{wdmw}; $0.35^{+0.50}_{-0.30}$ \cite{bruno}; and $-4.0$ \cite{hyc}.

The form factor $C(z,\nu)$ occurs via the pole diagrams of
Fig.~\ref{f: kpgep}.
The transition $K^+ \ra \pi^+ \pi^0$ occurs on-shell via the
$(27_L,1_R)$ operators only. The $K^+ \ra \pi^+ \pi^0$ decay, followed by
$\pi^0 \ra \gamma \gamma$ gives a very large contribution to
$K^+ \ra \pi^+ \gamma \gamma$, however, it can be subtracted
by implementing a cut in the invariant mass of the photon pair to
exclude the region near $m_{\pi^0}$. The off-shell transition is mediated
by the octet operator in Eq.~\ref{wlt} and results in \cite{eprt}:
\beq
C(z,\nu)={\alpha \over \pi} G_8\biggl[\frac{z-r^2_\pi}{z-r^2_\pi
+ir_\pi\Gamma_{\pi^0}/m_K}-\frac{z-{1 \over 3}(2+r^2_\pi)}
{z-r^2_\eta}\biggr]
\label{cffcha}
\eeq
The two form factors do not interfere, and one obtains:
\beq
{d\Gamma \over dz}={m^5_K \over 1024 \pi^3}z^2\lambda^{1\over 2}
(1,z,r^2_\pi)\biggl(|A(z)|^2+|C(z)|^2\biggr)
\label{ratez}
\eeq
In this case, $\chi$PT predicts
a correlation between the rate and the spectrum. The rate is given by the
expression:
\beq
B(K^+ \ra \pi^+ \gamma \gamma) = (5.26 + 1.64 \hat{c} + 0.32 \hat{c}^2
+ 0.49)\times 10 ^{-7}.
\label{ratekp}
\eeq
and the current experimental limit \cite{pgg787} is:
\beq
B(K^+ \ra \pi^+ \gamma \gamma) < \cases{
1.0 \times 10^{-6} & Phase space spectrum \cr
1.5 \times 10^{-4} & $\chi$PT spectrum \cr}
\label{exlimkp}
\eeq
The rate for $K^- \ra \pi^- \gamma \gamma$ can be obtained from
Eq.~\ref{kpppgg} by replacing $G_8$ and $\hat{c}$ with their complex
conjugates. An imaginary part of $\hat{c}$ would interfere with
the absorptive part of the amplitude (due to a real $\pi^+ \pi^-$
intermediate state) generating a CP odd rate asymmetry \cite{eprt}:
\beq
\Gamma(K^+ \ra \pi^+ \gamma \gamma) - \Gamma(K^- \ra \pi^- \gamma \gamma)
= 1.5 \times 10^{-23} {\rm Im}\hat{c}~{\rm GeV}
\label{cpaskpgg}
\eeq
To estimate ${\rm Im}\hat{c}$, the authors of Ref.~\cite{eprt} point out
that at the quark level, the CP phase appears via the electromagnetic
penguin operator \cite{gilwise}:
\beq
{\cal L} = -{G_F \over \sqrt{2}}|V_{ud}V_{us}|C_7 (\mu^2)\alpha
\overline{s}\gamma_\mu(1+\gamma_5 )d \overline{\ell} \gamma^\mu \ell
\label{empen}
\eeq
This operator transforms as an octet under $SU(3)_V$.
By requiring its chiral realization to
transform in the same way,
the authors of Ref.~\cite{eprt} conclude that:
\beq
{\rm Im}(g_8w_1)
\approx  {1\over 3 \pi^2} {\rm ln}\biggl({m_t \over m_c}\biggr)
A^2 \lambda^4 \eta
\label{imwg}
\eeq
Arguing that this is the only contribution to the phase of $\hat{c}$
they find:
\beq
|{\rm Im}\hat{c}| = {32 \over 3} \pi^2 |{\rm Im}w_1| \approx
0.005 \eta
\eeq
Noticing that the rate for $K^+ \ra \pi^+ \gamma \gamma$ given by
Eq.~\ref{ratekp} has a minimum for $\hat{c}=-2.56$, they conclude that:
\beq
\biggl|
{\Gamma(K^+ \ra \pi^+ \gamma \gamma)-\Gamma(K^- \ra \pi^- \gamma \gamma)
\over
\Gamma(K^+ \ra \pi^+ \gamma \gamma)+\Gamma(K^- \ra \pi^- \gamma \gamma)}
\biggr| \leq 0.002 \eta
\label{largeasy}
\eeq
However, the authors of Ref.~\cite{bruno} have pointed out that the
electromagnetic penguin operator also contributes a phase to $w_4$. They
find that this contribution essentially cancels the one from $w_1$ so that
${\rm Im}\hat{c} =0$ and there is no charge asymmetry in the standard
model. Given the potentially large asymmetry, Eq.~\ref{largeasy}, it
is important to resolve this issue.

\subsection{$K_L \ra \gamma \gamma$}

This process is not a very rare decay, but its calculation in $\chi$PT
illustrates some of the problems that one encounters in other
weak decays. To order ${\cal O}(p^4)$ this process occurs via the
$\pi^0,\eta$ poles of Fig.~\ref{f: akgg}.
Using Eqs.~\ref{wzwl} and \ref{wlt} one
finds in the limit of CP conservation that
$M(K_L\ra \gamma(q_1)\gamma(q_2))$ is:
\beq
M={2m^2_Kf^2_\pi \over m^2_K -m^2_\pi}{\alpha G_8\over \pi f_\pi}
\biggl(1+{1\over3}{m^2_K -m^2_\pi \over m^2_K - m^2_\eta}\biggr)
\epsilon_{\mu\nu\rho\sigma}q_1^\mu q_2^\nu \epsilon_1^\rho \epsilon_2^\sigma
\label{klga}
\eeq
At this order in $\chi$PT, however, we are instructed to use the
Gell-Mann-Okubo mass relation Eq.~\ref{gmor}, so the amplitude
vanishes. The first non-zero contributions result from higher order
$SU(3)$ breaking. We can parameterize these higher order effects
with a constant $F_1$, such that:
\beq
M={\alpha \over \pi} 2G_8f_\pi F_1
\epsilon_{\mu\nu\rho\sigma}q_1^\mu q_2^\nu \epsilon_1^\rho \epsilon_2^\sigma
\label{klfd}
\eeq
to find:
\beq
B(K_L \ra \gamma \gamma) =\cases{
(7.3 \times 10^{-4})|F_1|^2 &  \cr
(5.70 \pm 0.27)\times 10^{-4} & PDB average \cite{pdb} \cr}
\eeq
A fit to the experimental result yields a value $|F_1|\approx 0.88$.
A detailed analysis that tried to predict the value of $F_1$ was carried out
in Ref.~\cite{lin}. It included $SU(3)$ breaking effects in the vertices
as well as the decay constants; and also the effects of the $SU(3)$ singlet and
of $\eta-\eta^\prime$ mixing.
Ref.~\cite{lin} found that it was possible to accommodate
the experimental result, but
that it was not possible to predict it with any certainty.
Recently, the
authors of Ref.~\cite{vene} have criticized the treatment of the $SU(3)$
singlet in Ref.~\cite{lin}, however, this does not change the conclusion.

It is also of interest to study this amplitude for off-shell photons,
since this will give the dominant contribution to the decays
$K_L \ra \ell^+ \ell^- \gamma$
and $K_L \ra \ell^+ \ell^- \ell^+ \ell^-$.
In the limit of CP conservation, gauge
invariance requires the amplitude for $K_L \ra \gamma^* \gamma^*$ to be
of the form:
\beq
M= M(K_L \ra \gamma(q_1) \gamma(q_2))C(q_1^2,q_2^2)
\label{klgsgs}
\eeq
where we have normalized it so that $C(0,0)=1$. Bose symmetry requires
$C(q_1^2,q_2^2)$ to be symmetric under $q_1 \leftrightarrow q_2$, so we
write:
\beq
C(q_1^2,q_2^2)=1 + \sigma \biggl({q_1^2 +q_2^2 \over m^2_K}\biggr)+\cdots
\label{kgsgsff}
\eeq

\subsection{$K_L \ra \gamma \gamma$ Dalitz decays}
\label{s: dalitz}

Using the previous result, and defining $r_\ell = m_\ell  /m_K$,
$z=m_{\ell\ell}^2/m^2_K$, we can now study the process
$K_L \ra \ell^+ \ell^- \gamma$. The differential decay rate is:
\beq
{d\Gamma \over dz} = \Gamma(K_L \ra \gamma \gamma){2\alpha \over 3 \pi}
\biggl|C(0,z)\biggr|^2{(1-z)^3 \over z}\sqrt{1-{4 r^2_\ell \over z}}
\biggl(1+2{r^2_\ell \over z}\biggr),
\label{kllgdif}
\eeq
and the total rates are:
\beq
{\Gamma(K_L \ra \ell^+ \ell^- \gamma)\over \Gamma(K_L \ra \gamma
\gamma)}=\cases{
0.016 (1 + 0.05 \sigma) & $\ell=e$ \cr
0.016 \pm 0.001 & AGS-845, NA31 \cite{eeg845,eegNA31} \cr
4.09 \times 10^{-4}(1+0.64\sigma) & $\ell=\mu$ \cr
(6.81 \pm 0.64) \times 10^{-4} & FNAL-799 \cite{mdexp} \cr}
\label{kllgres}
\eeq
It has become usual in the literature to parameterize the form factor
following a model of Bergstrom {\it et. al.} \cite{berg},
in terms of a constant $\alpha_K$. This model, with $\alpha_K=0$ corresponds
to the vector dominance model of Quigg and Jackson \cite{quijak}.
Expanding the pole model parameterization, we identify
$\sigma= 0.418-1.29\alpha_K$. In terms of this pole model, AGS-845
\cite{eeg845,eegNA31} found
$\alpha_K = -0.28 \pm 0.083$ by fitting the $m_{\ell\ell}$
spectrum with QED radiative
corrections, and $\alpha_K = -0.18\pm 0.077$ without
QED radiative corrections. This is consistent with the NA31 result,
$\alpha_K = -0.28\pm 0.13$.

The branching ratio $\Gamma(K_L \ra e^+ e^- \gamma)/\Gamma(K_L \ra \gamma
\gamma)$ is not very sensitive to the form factor. With the form
Eq.~\ref{kgsgsff}, we find a variation from 0.0162 to 0.0165 when we take
$\alpha_K$ from $0$ to $-0.28$. With the pole model of Ref.~\cite{berg},
the branching ratio varies from 0.0163 to 0.0167 for this same range
of $\alpha_K$.

The branching ratio $\Gamma(K_L \ra \mu^+ \mu^- \gamma)/\Gamma(K_L \ra \gamma
\gamma)$, is more sensitive to the form factor. With the same range for
$\alpha_K$ as before, it varies from $(5.18\; {\rm to}\; 6.13) \times 10^{-4}$
using Eq.~\ref{kgsgsff}; and from $(5.58 \;{\rm to}\;7.28)\times 10^{-4}$
with the model of Ref.~\cite{berg}. The preliminary result from FNAL-799
cited in Eq.~\ref{kllgres}, can be used to extract from Ref.~\cite{berg}:
\beq
\alpha_K  =  -0.21\pm 0.11
\label{alphaberg}
\eeq
Which is consistent with the value of $\alpha_K$ found by the $K_L \to
e^+ e^- \gamma$ experiments, and inconsistent with $\alpha_K = 0$ at
the two $\sigma$ level.  Preliminary reports from FNAL-799 indicate
that the $m_{\ell\ell}$ spectrum for this decay is adequately fit by an
unmodified Kroll-Wada form \cite{krollw}.

	Finally, there have been several recent results on the
double Dalitz decay,
$K_L \ra e^+ e^- e^+ e^-$\cite{dedexp,eeee845,eeee137,eeee799},
which are summarized in Table~\ref{t: K4e}.

\begin{table}[htb]
\centering
\caption{Experimental Results on $K_L \to e^+ e^- e^+ e^-$}
\begin{tabular}{|l|c|c|} \hline
Experiment & Result &  Comments \\ \hline
NA31  & $ (4 \pm 3) \times 10^{-8}$  \cite{dedexp}&
the first 2 events \\ \hline
AGS-845   & $(3.04 \pm 1.24 \pm 0.26) \times 10^{-8}$ \cite{eeee845}&
6 events \\ \hline
KEK-137  & $(6 \pm 2 \pm 1) \times 10^{-8} $ \cite{eeee137}& partial
reconstruction \\ \hline
FNAL-799  & $(4.17 \pm 0.83) \times 10^{-8} $ \cite{eeee799} & 28 events,
preliminary result \\
\hline
\end{tabular}
\label{t: K4e}
\end{table}

	These results should be compared with the theoretical
expectation of $3.6 \times 10^{-8}$ ~\cite{quijak}.  Larger samples
will allow form factor effects to be studied.  FNAL-799 also expects
to observe the closely related process $K_L \to \mu^+ \mu^- e^+ e^-$
which is predicted at $0.8 \times 10^{-9}$ \cite{miyaz}.
Reference \cite{miyaz} also predicts $B(K_L \to \mu^+ \mu^- \mu^+ \mu^-)
= 5.4 \times 10^{-13}$, putting it beyond reach for the moment.
The predictions for these modes that one finds in the literature do not
always agree \cite{quijak,miyaz}.

\subsection{Direct emission $K_L \ra \pi^+ \pi^- \gamma$}

In the limit of CP conservation, this process does not occur
at order ${\cal O}(p^2)$ \cite{oldl,coska}.
The lowest order weak Lagrangian Eq.~\ref{wlt}
contributes to the process $K^0_1 \ra \pi^+\pi^-\gamma$, and thus it
contributes to $K_L \ra \pi^+ \pi^- \gamma$ an indirect CP violating term.
This term has a characteristic bremsstrahlung spectrum and has been
observed \cite{carroll,kppgfnal}.
Of greater interest to us is the so called direct emission term.
{}From a theoretical point of view, we will define it as the amplitude that
starts at ${\cal O}(p^4)$. Experimentally, it is observed by subtracting
the bremsstrahlung portion from the full amplitude. Gauge invariance
requires the amplitude for $K_L(k)\ra \pi^+(p^+)\pi^-(p^-)\gamma(q)$
to be of the form:
\beq
M={eG_8 f^2_\pi \over m_K} \epsilon^\mu \biggl[
\xi_E(z,\nu)\biggl(\nu (p^+ + p^-)_\mu - z(p^+-p^-)_\mu \biggr)
+{4 i \over m^2_K} \xi_M(z,\nu) \epsilon_{\mu\nu\alpha\beta}k^\nu p^{+\alpha}
p^{-\beta} \biggr]
\label{gikppg}
\eeq
where $z=2k\cdot q / m^2_K$, $x^\pm =2k \cdot p^\pm /m^2_K$, and
$\nu=x^+-x^-$.
The ``magnetic'' form factor $\xi_M$ receives contributions from $\pi^0,\eta$
poles as in Fig.~\ref{f: mkppg}, and vertices from Eq.~\ref{wzwl}. At order
${\cal O}(p^4)$ we find:
\beq
\xi_M = {m^2_K \over m^2_K - m^2_\pi}{m^3_K \over 8 \pi^2 f^3_\pi}
\biggl(1+{1\over 3}{m^2_K -m^2_\pi \over m^2_K - m^2_\eta}\biggr)
\label{magwz}
\eeq
However, when the Gell-Mann-Okubo relation (Eq.~\ref{gmor})
is applied, one gets
the same cancellation that occurred in
$K_L \ra \gamma \gamma$.
We can parameterize the $SU(3)_V$ violating terms that make the
amplitude non-zero, as well as terms of order ${\cal O}(p^6)$ with
a naive expansion of the form factors. We follow Ref.~\cite{coska} and
include the strong rescattering phases $\delta^I_J$ for the spin $J$,
isospin $I$ $\pi-\pi$ scattering at a center of mass energy squared
$(p_+ + p_-)^2 = m^2_K(1-z)$. With all this we write:
\beqn
\xi_E(z,\nu) &=& i{8(m_K^2-m^2_\pi)\over m_K f_\pi}{\eta_{+-}\over z^2-\nu^2}
+\biggl[F_E\biggl(1+\sigma_E z\biggr)e^{i(\delta^1_1 - \delta^0_0)}
+ig_E \nu e^{i(\delta^0_2 -\delta^0_0)} \biggr]  \nonumber\\
\xi_M(z,\nu) &=& {m^3_K \over 8 \pi^2 f^3_\pi}
\biggl[F_M\biggl(1+\sigma_M z\biggr)e^{i(\delta^1_1 - \delta^0_0)}
+ig_M \nu e^{i(\delta^0_2 -\delta^0_0)} \biggr].
\label{magfd}
\eeqn
We have included in $\xi_E$ the inner bremsstrahlung contribution which can
interfere with $F_E$ in the rate. We have not included possible ${\cal O}(p^6)$
$\Delta I = 3/2$ terms. The terms $F_E$,
and $g_M$ are CP violating. The differential decay rate is given by:
\beq
{d\Gamma \over dz d\nu}={\alpha m_K \over 128 \pi^2}G_8^2 f^4_\pi
\biggl(|\xi_E|^2 + |\xi_M|^2\biggr)
\biggl[(1-z)(z^2-\nu^2)-4r^2_\pi z^2\biggr]
\label{kpggdr}
\eeq
A detailed analysis of the Dalitz plot for this decay should allow a
determination of the parameters in Eq.~\ref{magfd}.
The simplest thing we can do is to set all the new constants in
Eq.~\ref{magfd} to zero except $F_M$. In that case
we find for the direct emission (with $E_\gamma > 20~{\rm MeV}$):
\beq
B(K_L \ra \pi^+ \pi^- \gamma)_{DE} =\cases{
(3.35 \times 10^{-5})|F_M|^2 &  \cr
(2.89 \pm 0.28)\times 10^{-5} & BNL \cite{carroll} \cr
(3.19 \pm 0.16)\times 10^{-5} & FNAL-731 \cite{kppgfnal} \cr}
\label{kppgmf}
\eeq
which results in $F_M \approx 0.95$. One can try to predict $F_M$ in
a model that includes $SU(3)$ breaking effects, similar to
the one used by Ref.~\cite{lin} for $K_L \ra \gamma \gamma$. One
finds that it is possible to
accommodate the result \cite{linv}, but not to predict it with certainty.
If we set $F_M$ to zero and keep $F_E$ in Eq.\ref{magfd}, we find that
this direct
emission rate is also consistent with an interference of the bremsstrahlung
amplitude and a CP violating term $F_E \sim 1.4$.
Keeping only $F_M$ and $F_E$, Ref.~\cite{carroll} found
a best fit to the spectrum with an admixture
of CP violating direct decay $F_E \approx 0.26 F_M$,
lying $< 2 \sigma$ from the no-interference fit.
However, naive dimensional
analysis suggests that CP violation in the standard model
is much smaller than this \cite{linv}. The fits of Ref.~\cite{carroll}
found no evidence for the quadrupole term $g_E$ at the
$25\%$ level.

Assuming CP conservation in the direct emission,
a fit to the distribution $d\Gamma /dz$
provides information on the slope $\sigma_M$:
\beq
\sigma_M = \cases{-0.91 & FNAL-731 \cr
           -0.89   & BNL \cr}
\label{kpggmg}
\eeq
One can resort to models to predict the ${\cal O}(p^6)$ form factors like
$\sigma_M$ \cite{linv,piccio,enp}. It appears that a vector meson
dominance model does not reproduce the spectrum shape unless one
includes $SU(3)_V$ breaking effects. A more careful study is needed.

By observing the interference between $K_S$ and $K_L$ decays into
$\pi^+ \pi^- \gamma$, FNAL-731 \cite{kppgfnal} has measured:
\beq
|\eta_{+-\gamma}|={M(K_L \ra \pi^+ \pi^- \gamma)_{E1} \over
M(K_S \ra \pi^+ \pi^- \gamma)_{E1}} = (2.15 \pm 0.26 \pm 0.20)
\times 10^{-3}
\eeq
with a phase $\phi_{+-\gamma}=(72 \pm 23 \pm 17)^\circ$. Since
$|\eta_{+-}| = (2.268 \pm 0.023)\times 10^{-3}$ \cite{pdb}, the FNAL-731
result is consistent with no CP violation beyond that present in
$K \ra \pi \pi$ decays.

Additional information can be obtained by
studying the Dalitz decay $K_L \ra \pi^+ \pi^- e^+ e^-$. Separating
the I.B and D.E. contributions, and using for the D.E. a constant
coupling $F_M$ that fits the $(K_L \ra \pi^+ \pi^- \gamma)_{DE}$ rate,
Ref.~\cite{manf}
calculates $B(K_L \ra \pi^+ \pi^- e^+ e^-)= (1.3 + 1.8) \times 10^{-7}$.
This decay has been recently observed by FNAL-799, but a branching ratio
has not yet been reported.
%$B(K_L \ra \pi^+ \pi^- e^+ e^-) = ????$.
The authors
of Ref.~\cite{manf} have also studied a CP odd
correlation between the $\pi^+ \pi^-$ and $e^+ e^-$
planes and predict a $14\%$ asymmetry. Unfortunately
this asymmetry is mostly due to indirect CP violation through the parameter
$\epsilon$, making an extraction of direct CP violation very difficult.

\clearpage

\section{Decays into charged lepton pairs}
\label{s: kll}

These modes generally receive both long and short distance contributions
that are of comparable size. Some of them are dominated by the long
distance component, but exhibit interesting interference effects with the
smaller short distance contribution.
Their primary interest is the study of the short distance parameters
$\rho$ and $\eta$, but to do so one needs some understanding of the long
distance contributions. In this respect they also allow one to test the
ideas of $\chi$PT. The observables that probe the short distance
physics are also sensitive to new interactions, however,
we will restrict ourselves to a discussion of  the standard model.

\subsection{Short Distance $K_L \ra \ell^+ \ell^-$}

The short distance contribution to these processes comes from the
box and electro-weak penguin diagrams of Fig.~\ref{f: sdll}.
To compute the amplitude for this process, one
relates the matrix element $<0|\overline{s}\gamma_\mu
(1+\gamma_5 )d|K_L>$ to that occurring in $K^+ \ra \ell^+ \nu$
(see Eq.~\ref{curbos}). The top-quark contribution is easily
computed to be \cite{inami}:
\beq
{B(K_L \ra \ell^+ \ell^-)_{SD} \over B(K^+ \ra \ell^+ \nu)}
= {\tau_{K_L} \over \tau_{K^+}}
\biggl({\alpha \over \pi \sin^2\theta_W}\biggr)^2 A^4 \lambda^8
(1-\rho)^2 \biggl[ Y(x_t) \biggr]^2
\label{sdkll}
\eeq
where the function $Y(x_t)$ is given by:
\beq
Y(x_t)={x_t \over 8}\biggl({x_t-4\over x_t-1}+{3x_t \over (x_t-1)^2}{\rm ln}
x_t\biggr)
\label{yxt}
\eeq
An approximate expression is provided by Ref.~\cite{buras}
$Y(x_t) \approx 0.315 x_t^{0.78}$. The contribution of the charm-quark,
with perturbative QCD corrections can be found in Ref.~\cite{burasb}.
These authors provide us with the approximate expression for the
complete result:
\beq
B(K_L \ra \mu^+ \mu^-)_{SD} = 1.7\times 10^{-10}  x_t^{1.56} A^4
(\rho_0 -\rho)^2
\label{kmmbu}
\eeq
where deviations of $\rho_0$  from 1 measure the
charm-quark contribution with QCD corrections. For typical values of all the
parameters involved, $\rho_0 \sim 1.27$.

An interesting feature of this decay is the longitudinal polarization
of the final lepton, a CP violating observable \cite{herczeg}:
\beq
P_L \equiv {N_L - N_R \over N_L + N_R}
\label{kllasy}
\eeq
The latest estimate of this quantity within the standard model is
about $2 \times 10^{-3}$ \cite{epmupo}. This number, however, is
directly proportional to $\epsilon$ \cite{botlim}. Since this is
probably too small to be measured, $P_L$ is a very good place to look
for direct CP violation outside the standard model.

\subsection{Long distance $K_L \ra \ell^+ \ell^-$}

The long distance contribution to these decays is expected
to be dominated by the two photon intermediate state as in
Fig.~\ref{f: ldll}.
It is straightforward to compute the absorptive part of the amplitude
by using the experimental rate for $K_L \ra \gamma \gamma$.
The result is ($r^2_\ell = m^2_\ell / m^2_K$):
\beqn
B(K_L \ra \ell^+ \ell^-)_{abs} &=& {1 \over 2}\alpha^2 r^2_\ell
{1 \over \beta} \biggl|{\rm ln}{1+\beta \over 1-\beta}\biggr|^2
B(K_L \ra \gamma \gamma) \nonumber \\
\beta &=& \sqrt{1-4r^2_\ell}
\label{kmmab}
\eeqn
This can be compared to the latest measurements:
\beqn
B(K_L \ra \mu^+ \mu^-) &=& \cases{
(6.8 \pm 0.3) \times 10^{-9} & absorptive \cr
(7.9 \pm 0.7)\times 10^{-9} & KEK 137 \cite{mumu137} \cr
(6.86 \pm 0.37)\times 10^{-9} & AGS-791 \cite{ekmm} \cr} \nonumber \\
B(K_L \ra e^+ e^-) &= & \cases{
(3.0 \pm 0.1) \times 10^{-12} & absorptive \cr
< 4.7 \times 10^{-11} & AGS-791 \cite{mue791} \cr}
\eeqn
A calculation of the dispersive  part of the long distance effects
is not possible at present within $\chi$PT. There are two problems: the
on-shell $K_L \ra \gamma \gamma$ vertex cannot be computed reliably as
already explained; and the divergence of the loop diagram implies that
there are counterterms for this process. These counterterms are similar to
those occurring in the decay $\pi^0 \ra e^+ e^-$ \cite{simipi}, but they
are not the same and have not been
determined. However, we can resort to models to estimate the dispersive
contribution. A vector dominance model  gives results
that can be written compactly in the limit $m_\ell /m_P \ra 0$ \cite{quigg}:
\beqn
{\Gamma(P^0 \ra \ell^+ \ell^-) \over \Gamma(P^0 \ra \gamma \gamma)} &=&
{2 \alpha^2 m_\ell ^2 \over m_P^2}
\biggl[ X^2 + \biggl({\rm ln}{m_P \over m_\ell}\biggr)^2 \biggr] \nonumber \\
X&=& {\pi \over 12}+{1 \over 4\pi}+{1\over \pi}\biggl({\rm ln}{m_P \over
m_\ell }\biggr)^2-{3\over \pi}{\rm ln}{m_V \over m_\ell }
\label{quiggj}
\eeqn
where $m_V$ is the rho mass. Although this is a convenient expression for
decays into electrons, it is not accurate by about a factor of two for muons.
One also finds in the literature the model of Ref.~\cite{berg}, in which
one allows an additional form-factor for the $K_L \ra \gamma \gamma$ vertex
itself. This is modelled by a $K^* \ra \rho$ transition in terms of the
parameter $\alpha_K$ of section~\ref{s: dalitz}. The result can be written
as \cite{berg}:
\beq
B(K_L \ra \mu^+ \mu^-)^{\gamma\gamma}_{disp} =
4.7 \times 10^{-10} \biggl( 1.3 + 4.9 \alpha_K \biggr)^2,
\label{klldisp}
\eeq
and the $\alpha_K =0$ case corresponds to the vector dominance model of
Quigg and Jackson.
The contributions from a negative $\alpha_K$ tend to cancel the first
term in the above result, and this cancellation is almost complete
for $\alpha_K =-0.28$. We regard this cancellation as accidental,
and prefer the result with $\alpha_K =0$ as a more conservative
estimate for the long distance dispersive rate.

In view of the large model dependence in estimating the
long distance contribution to the dispersive part of
$K_L \ra \mu^+ \mu^-$, it would seem very
difficult to extract meaningful constraints on the short distance
contribution. Nevertheless, recent attempts to do
so can be found in Ref.~\cite{belgeng}.

{}From Eq.~\ref{quiggj} we can also compute the two-photon contribution to
$B(K_L \ra e^+ e^-)_{disp}$, we find
$4.6 \times 10^{-12}$. The long distance
contributions to $K_L \ra e^+ e^-$ are about 15 times as large as
one would expect by a naive scaling of $K_L \ra \mu^+ \mu^-$ with
$m^2_e / m^2_\mu$. This factor can be traced back to the logarithms in
Eqs.~\ref{kmmab},~\ref{quiggj}. There is no analogous factor in the
short distance contribution, which then has the naive scaling.
This means that it is even harder to observe the short
distance contribution in $K_L \ra e^+ e^-$ than it is in
$K_L \ra \mu^+ \mu^-$ \cite{rrw}.

\subsection{$K \ra \pi \gamma^*$}
\label{s: kpg}

This is the most important long distance contribution to
the CP conserving decays
of the form $K \ra \pi \ell^+ \ell^-$. It can be computed to ${\cal O}(p^4)$ in
$\chi$PT in terms of a few unknown constants.
The most general form allowed by electromagnetic gauge
invariance for the amplitude $A(K^{\pm,0}(k)\ra \pi^{\pm,0}(p)
\gamma^*(\epsilon,q))$ is \cite{rafao}:
\beq
A^{\pm,0}= {e G_8 \over 16 \pi^2}
\epsilon^\mu\biggl(q^2(k+p)_\mu -(m^2_K-m^2_\pi)q_\mu\biggr)
C^{\pm,0}(z)
\label{kpggi}
\eeq
where $z=q^2/m^2_K$.

It can be seen immediately that gauge invariance requires the amplitude
to have at least three external momenta (and there is one external photon),
so that it can only start at ${\cal O}(p^4)$.
Eq.~\ref{kpggi} also shows that the process
$K \ra \pi \gamma$ with an on-shell photon is forbidden by gauge invariance.
Computing the diagrams in Fig.\ref{f: kpgee}, one finds
the form factor $C(z)$ to
${\cal O}(p^4)$ in $\chi$PT in terms of the couplings of section~\ref{s: cpt}
and the function:
\beq
\phi_K(z) = \cases{ {1 \over 3} \biggl[ \biggl( {4 \over z} -1\biggr)^{3\over
2}
{\rm atan}\sqrt{{z \over 4-z}}
-{4 \over z} +{5\over 6}
\biggr] & $ z\leq 4$ \cr
{1 \over 3} \biggl[{1\over 4} \biggl( 1-{4 \over z}\biggr)^{3\over 2}
\biggl({\rm ln}\biggl({z-2-\sqrt{z(z-4)} \over
z-2+\sqrt{z(z-4)}  }\biggr) +i 2 \pi \biggr) -{4 \over z} +{5\over 6}
\biggr] & $ z > 4$ \cr}
\label{fphi}
\eeq
The function $\phi_\pi(z)$ is obtained by replacing
$z \rightarrow zm^2_K/m^2_\pi$
in Eq.~\ref{fphi}. One finds for the charged mode \cite{rafao}:
\beqn
C^\pm (z) &=& {1 \over 3}\biggl[(4\pi)^2 \biggl(w^r_1(\mu)-w^r_2(\mu)\biggr) +
3(4\pi)^2\biggl(w^r_2(\mu)-4L^r_9(\mu)\biggr) \nonumber \\
&+&{\rm ln}\biggl({m_K m_\pi \over
\mu^2} \biggr) \biggr] -\biggl(\phi_\pi(z)+\phi_K(z)\biggr) \nonumber \\
&\equiv & -\biggl(w_+ +\phi_\pi(z)+\phi_K(z)\biggr) ,
\label{cpm}
\eeqn
and for the neutral mode:
\beqn
C^0 (z) &=& -{\sqrt{2} \over 6}\biggl[(4\pi)^2
\biggl(w^r_1(\mu)-w^r_2(\mu)\biggr)+
{\rm ln}\biggl({m^2_K \over \mu^2}\biggr)\biggr]
+\sqrt{2} \phi_K \nonumber \\
&\equiv & \sqrt{2}\biggl({w_S \over 2} + \phi_K(z) \biggr)
\label{coo}
\eeqn

\subsection{$K^+ \ra \pi^+ \ell^+ \ell^-$}

We can now study the processes $K^+(k) \ra \pi^+(p) \ell^+(k^+) \ell^-(k^-)$
that are dominated
by the one photon intermediate state.  The matrix element is given
by ($z=(k-p)^2/m^2_K$):
\beq
M^{(1)}= -{\alpha G_8 \over 4 \pi}C^+(z)\overline{u}(k^-)
(\slash{k}+\slash{p})v(k^+)
\label{mwithl}
\eeq
{}From this, it is straightforward to compute the decay distribution
\cite{rafao}:
\beq
{d\Gamma \over dz} = {G_8^2 \alpha^2 m^5_K \over 12\pi (4\pi)^4}
\lambda^{3 \over 2}(1,z,r^2_\pi)
\biggl(1-4{r^2_\ell \over z}\biggr)^{1\over 2}
\biggl(1+2{r^2_\ell \over z}\biggr)
\biggl| C^+(z) \biggr|^2
\label{rwithl}
\eeq
This is shown in Fig.~\ref{f: kpllr}.
Integrating Eq.~\ref{rwithl} from $4r^2_\ell$ to $(1-r_\pi)^2$ and using
Eqs.~\ref{fphi},~\ref{cpm} we can get a prediction for the rates
$K^+ \rightarrow \pi^+ \ell^+ \ell^-$ in
terms of the unknown constant $w_+$:
\beqn
B(K^+\ra \pi^+ e^+ e^-) &=& (3.15-21.1w_+ + 36.1w^2_+)\times 10^{-8}
\nonumber \\
B(K^+\ra \pi^+ \mu^+ \mu^-) &=& (3.93-32.7w_+ + 70.5w^2_+)\times 10^{-9}
\label{kpllbr}
\eeqn
One can then determine the constant $w_+$ from measurements of the rate,
the spectrum, or both.   A measurement of $w_+$ in $K^+\ra \pi^+ e^+ e^-$
has recently been reported~\cite{alliegro} (AGS-777).
Eq.~\ref{wpres} and Fig.~\ref{f: zeller} give the result of a simultaneous
fit of the rate and spectrum shape:
\beqn
B(K^+ \rightarrow \pi^+ e^+ e^-) &= &
(2.99 \pm 0.22)\times 10^{-7} \nonumber \\
w_+ &=& 0.89^{+0.24}_{-0.14}
\label{wpres}
\eeqn
Eq.~\ref{kpllbr}, which has not been imposed on the fit, is shown as
a parabola in Fig.~\ref{f: zeller}.  It passes within about $1.5 \sigma$
of the best fit values.  As consistency checks, one can insert
these values in turn into Eq.~\ref{kpllbr} yielding:
\beqn
B(K^+ \rightarrow \pi^+ e^+ e^-) & = &
(1.30^{+1.24}_{-0.53})\times 10^{-7} \nonumber \\
w_+ &= &\cases{1.20\pm 0.033 & or \cr
 -0.62 \pm 0.033 & \cr}
\label{wpres2}
\eeqn
This approach to $K^+ \rightarrow \pi^+ e^+ e^-$ seems quite
promising, but more data is desirable.  A somewhat larger data
set with better systematics should be forthcoming from AGS-851.
In the longer term, a much larger sample is promised from AGS-865
which is presently under construction.

Some model calculations of $w_+$ have found: $0.7$ \cite{hyc};
$0.98^{+0.78}_{-0.39}$ \cite{bruno};
and $1.9$ \cite{wdmw}. If we take as a typical number
$w_+ = 0.89$ we then predict $B(K^+\ra\pi^+ \mu^+ \mu^-)=3.07\times 10^{-8}$.
At present there is only an upper limit
of $<2.3 \times 10^{-7}$ on this decay from AGS-787 ~\cite{pmm787},
but a data set
with sensitivity $< 10^{-8}/$~event is presently under analysis by
this experiment.

A CP violating imaginary part of the constant $w_+$ would interfere
with the absorptive part generated by the two pion intermediate state
(Eq.~\ref{fphi}), giving rise to a CP odd rate asymmetry \cite{eprt}:
\beq
\biggl|{\Gamma(K^+ \ra \pi^+ e^+ e^-)-\Gamma(K^- \ra \pi^- e^+ e^-) \over
\Gamma(K^+ \ra \pi^+ e^+ e^-)+\Gamma(K^- \ra \pi^- e^+ e^-)}\biggr|
\approx 0.01 {\rm Im}w_+ \approx 3 \times 10^{-5} \eta ,
\eeq
where in the last step we have used the estimate of Ref.~\cite{eprt},
${\rm Im}w_+ \approx 0.003 \eta$. This is too small to see in the near
future.

We should comment on a possible parity violating asymmetry for
these decays. It was pointed out in Ref.~\cite{savawi} that apart from the
one-photon intermediate state, these processes have a short distance
contribution from  a $Z$ intermediate
state or from box diagrams as in  Fig.~\ref{f: sdll}.
These new operators generate a second possible
amplitude in addition to Eq.~\ref{mwithl}:
\beq
M^{(2)} = {G_F \over \sqrt{2}}V_{us} \alpha \xi
\overline{u}(k^-)(\slash{k}+\slash{p}) \gamma_5 v(k^+)
\label{pvamp}
\eeq
where we have used the lowest order $\chi$PT result $f_+ =1$, $f_- =0$.
The constant $\xi$ contains the short distance factors,
it is given in Ref.~\cite{sawi}:
\beq
\xi = -1.4\times 10^{-4} - {Y(x_t) \over 2 \pi \sin^2\theta_W}
A^2 \lambda^4 (1-\rho - i \eta),
\label{expxi}
\eeq
where $Y(x_t)$ is given in Eq.~\ref{yxt}.
The first term in Eq.~\ref{expxi}
corresponds to the charm-quark contribution with QCD
corrections, and the second term to the top-quark contribution.
If we denote by $\Gamma_L$, $\Gamma_R$, the rates for producing a
left-handed (right-handed) $\mu^+$ in $K^+ \ra \pi^+ \mu^+ \mu^-$, then the
interference of the two amplitudes generates the parity
violating observable \cite{sawi}:
\beq
|\Delta_{LR}|=\biggl|{\Gamma_R -\Gamma_L \over \Gamma_R + \Gamma_L}\biggr|
\approx 2.3 {\rm Re}\xi
\eeq
A measurement of $|\Delta_{LR}|$ at the tenth of a percent level would provide
valuable information on the CKM parameter $\rho$ \cite{sawi,bg}.
Taking for example,
$m_t = 140~{\rm GeV}$, $\rho =-0.51$, Eq.~\ref{expxi} yields $|\Delta_{LR}|
= 3.7 \times 10^{-3}$.

Finally, the authors of Ref.~\cite{beng} have proposed some T-odd observables
that can be studied in this decay. Unfortunately, in order to extract
information on CP violation from this type of observable one must be able
to reliably subtract unitarity effects. This usually involves a comparison
of the two charge conjugated modes. In Ref.~\cite{beng} it is found that
the unitarity effects are small in asymmetries where the polarization of
both the $\mu^+$ and the $\mu^-$ are measured.   Unfortunately such
measurements are extremely difficult, due to the absorption of stopped
$\mu^-$ before they can decay.

\subsection{$K_1^0 \ra \pi^0 \ell^+ \ell^-$}

In the limit of CP conservation, the one-photon intermediate state
contributes only to $K_1^0(k) \ra \pi^0(p) \ell^+(k^+) \ell^-(k^-)$.
The decay distributions can be predicted in terms of the constant $w_S$ with
Eq.~\ref{rwithl} simply using:
\beq
C^0_1(z)=w_S+2\phi_K(z)
\label{kshpig}
\eeq
Ignoring CP violation this results in rates and spectra for the decays
$K_S \ra \pi^0 \ell^+ \ell^-$. For the rates one finds ~\cite{pmm787}:
\beqn
B(K_S \ra \pi^0 e^+ e^-) &=& (3.07-18.7w_S +28.4w^2_S)\times 10^{-10}
\nonumber \\
B(K_S \ra \pi^0 \mu^+ \mu^-) &=& (6.29-38.9w_S +60.1w^2_S)\times 10^{-11}
\label{ratekspll}
\eeqn
In general, $\chi$PT does  not
relate $w_S$ to $w_+$ measured in the charged mode. Without additional
input one must first measure $w_S$ in one of the decays and then use it
to predict the others. The prediction one finds in the literature is based
on an additional assumption. The authors of Ref~\cite{eprt} demanded
that the meson Lagrangian transform under $SU(3)_V$ as a pure octet in analogy
with the quark electromagnetic penguin operator. That gave them
the constraint $w_2 = 4 L_9$ which then allowed them to predict:
\beq
w_S = w_+ +{1\over 6}{\rm ln}\biggl({m^2_\pi \over m^2_K}\biggr)
\label{drpre}
\eeq
One must remember, however, that Eq.~\ref{drpre} is an assumption that
goes  beyond $\chi$PT, and that it is not satisfied in some models
\cite{bruno}. Sample model calculations of $w_S$ yield: $0.3$ \cite{hyc};
$0.98^{+0.98}_{-0.59}$ \cite{bruno}; and $1.4$ \cite{wdmw}.

\subsection{$K_L \rightarrow \pi^0 e^+ e^-$}
\label{s: klp0}

The decay $K_L \rightarrow \pi^0 e^+ e^-$ is significantly more complicated
than the others we have been discussing. It has at least three different
contributions that could be of the same size. The most interesting one,
of course, is the direct CP violation. It originates in the
diagrams of Fig.~\ref{f: skee}. The result has been
computed in Refs.~\cite{burasb,gilwise,otherpiee}. The full result has
a complicated form, but Ref.~\cite{buras} gives an approximate expression:
\beq
B(K_L \rightarrow \pi^0 e^+ e^-)=0.32 \times 10^{-10} \eta^2 A^4I(x_t)
\label{kpeebu}
\eeq
with $I(x_t) \approx 0.73 x_t^{1.18}$.
With present day bounds on all the parameters, Ref.~\cite{buras}
finds that this contribution to the rate ranges from about $10^{-12}$
to $2 \times 10^{-11}$.

There is also an indirect CP violating contribution, that is, one that
proceeds via the parameter $\epsilon$ in the mass matrix. Its contribution
cannot be computed directly at present, and its precise value will only be
known after a measurement of $K_S \rightarrow \pi^0 e^+ e^-$ is done. However,
one can also use Eq.~\ref{kshpig} to predict the indirect CP violation in the
decay $K_L \ra \pi^0 \ell^+ \ell^-$:
\beqn
B(K_L \ra \pi^0 \ell^+ \ell^-)_{ind} &=&
|\epsilon|^2{\tau_{K_L} \over \tau_{K_S}}B(K_S \ra \pi^0 \ell^+ \ell^-)
\nonumber \\
B(K_L \rightarrow \pi^0 e^+ e^-)_{ind} &< & 1.27 \times 10^{-12}
\label{indcp}
\eeqn
The last result follows from using $w_+ = 0.89^{+0.24}_{-0.14}$,
and Eq.~\ref{drpre}
to obtain $w_S = 0.47^{+0.24}_{-0.14}$. With this range for
$w_S$, the rate Eq.~\ref{indcp} varies by more than three
orders of magnitude! In fact, for $w_S = 0.33$, Eq.~\ref{ratekspll} is
not sufficiently accurate to calculate the rate.
Given this large sensitivity to
$w_S$, and the fact that we rely on Eq.~\ref{drpre}, this
result must be viewed with caution.

Finally, we can use the result for $K_L \ra \pi^0 \gamma \gamma$,
Eq.~\ref{pocl}, to estimate the rate for the
CP conserving part of the $K_L \rightarrow \pi^0 e^+ e^-$ amplitude. As
usual \cite{seg},
we will simply give the contribution from the absorptive part
of the two photon intermediate state, depicted in Fig.~\ref{f: ldee}.
The contribution from $A(z,\nu)$
is suppressed by $m_e$ and can be neglected.
Using Eq. \ref{pocl}~ we find a simple result if $B(z,\nu)$ is constant
or if it depends only on $z$. We find \cite{bdv}:
\beq
A_{abs}(K_L(p) \rightarrow \pi^0 e^+(k^{\prime}) e^-(k)) = -\frac{2\alpha^2}
{3\pi} a_V{G_8 \over m^2_K}
p\cdot (k -k^{\prime}) \overline{u} \slash{p} v
\label{kpeeab}
\eeq
After squaring and integrating over phase space, this gives a lower limit for
the branching ratio from the CP conserving amplitude.
With $-0.32 < a_V < 0.19$, Eq. \ref{avbou}, Ref.~\cite{pggNA31} quotes:
\beq
B_{CP}(K_L \rightarrow \pi^0 e^+ e^-) \leq 4.5 \times 10^{-13}
\label{rate}
\eeq
However, we must remember that this is only the absorptive part
of the amplitude, so that the number is not really an upper bound.
However, it is sufficiently smaller than the direct CP-violating
component, that it will probably not impede efforts to extract
the latter.

The fact that the direct CP-violating contribution to this
decay is comparable or greater than the competing contributions,
has sparked a good deal of experimental interest, including
a number of dedicated searches.
The present situation is summarized in Table~\ref{t: kpiee}.

\begin{table}[htb]
\centering
\caption{Summary of $K_L \to \pi^0 e^+ e^-$ Experiments}
\begin{tabular}{|l|c|c|c|} \hline
Experiment & Result & Status & Comments \\ \hline
NA31 & $ <4 \times 10^{-8}$ \cite{peeNA31}  & finished &
 \\ \hline
FNAL-731  & $ < 7.5 \times 10^{-9} $ \cite{pee731} & finished& \\ \hline
AGS-845  & $ < 5.5 \times 10^{-9} $ \cite{kpee845} & finished &\\ \hline
FNAL-799 &   & analyzing& aims for $10^{-10} - 10^{-9}$ sensitivity \\ \hline
KEK-162 &    & running &aims for $10^{-10}$ sensitivity \\ \hline
FNAL-799II &   & under construction & aims for $< 10^{-10}$ sensitivity \\
\hline
\end{tabular}
\label{t: kpiee}
\end{table}

Although this decay has a good  kinematic signature and
its all-electromagnetic final state can be exploited in
the design of experiments,
it has unfortunately been found to suffer from the complication of a
very difficult background.  This stems
from the processes shown in Fig.~\ref{f: bkee}, i.e.
radiative corrections to $K_L \to e^+ e^- \gamma$,
resulting in $K_L \to e^+ e^- \gamma \gamma$.   This was first observed
by AGS-845~\cite{eegg845} in the course of their search for
$K_L \to \pi^0 e^+ e^-$~\cite{kpee845}.   The branching ratio was
measured to be $(6.6 \pm 3.2) \times 10^{-7}$.   The potency of this
process as a background to $K_L \to \pi^0 e^+ e^-$, which had not
previously been appreciated, was explicated by Greenlee~\cite{green}.  Although
there is no particular enhancement near $m_{\gamma \gamma} =
m_{\pi^0}$, he found that the rate is sufficient to make the
extraction of a $K_L \to \pi^0 e^+ e^-$ signal at the $10^{-11}$ level
problematic.  For example, assuming an acceptance interval $\Delta
m_{\gamma \gamma} = 5$ {\rm MeV}, with highly optimized kinematic cuts that
accept half the $K_L \to \pi^0 e^+ e^-$ events, the background enters
at an equivalent branching ratio of $10^{-10}$.  If the cuts are
further tightened to include only $10 \%$ of the signal events, the
background is reduced by only about a factor $3$.  Further progress
can be made if the resolution in $m_{\gamma \gamma}$ can be improved
beyond the already very optimistic assumption used by Greenlee.
Another approach to coping with this background is to attempt to subtract
it, capitalizing on its smooth dependence upon $m_{\gamma \gamma}$.  Of
course this requires high statistics, which are not easy to come by at
the required level of sensitivity.

\subsection{$K_L \ra \pi^0 \mu^+ \mu^-$}

This decay mode, unlike the previous one, receives a substantial contribution
from the two photon amplitude at ${\cal O}(p^4)$ since the muon mass is not
negligible. This results in the possibility of substantial interference
between the CP conserving and violating amplitudes.
The amplitude can be written as:
\beq
M = {\alpha \over 4 \pi}{\rm Re}G_8\overline{u}(k^-)[im_\mu h(z) -
(\slash{k} + \slash{p})g(z)]v(k^+)
\eeq
The CP violating form factor is
\beq
g(z)=\epsilon[{\rm Re}w_S + 2 \phi_K (z)] + i {\rm Im}w_S
\eeq
whereas the CP conserving form factor is given by \cite{eprt}:
\beq
h(z)={\alpha \over \beta z}{\rm ln}\biggl({1-\beta \over 1 + \beta}\biggr)
\biggl[(z-r^2_\pi)F\biggl({z \over r^2_\pi} \biggr) -(z-1-r^2_\pi)F(z)\biggr]
\eeq
where $\beta = \sqrt{1-4r^2_\mu/z}$.
An interference between the two amplitudes generates a CP violating
muon polarization. Taking for example ${\rm Re}w_S = 0.73$ and
${\rm Im}w_S=0.001$ Ref.~\cite{eprt} finds an average transverse muon
polarization $<\xi> = -0.37$ and a branching ratio
$B(K_L \ra \pi^0 \mu^+ \mu^-)=6.3 \times 10^{-12}$. At present there
is a preliminary limit $B(K_L \ra \pi^0 \mu^+ \mu^-)<1.7 \times 10^{-8}$ from
FNAL-799 \cite{mdexp}.

\clearpage

\section{The experiments}
\label{s: exp}

	In this section  recent and current
experiments are described and their results summarized.
We exclude  those results not directly related to the subject
of this review (e.g. $\epsilon'/\epsilon$, rare $\pi$ decay).
We also discuss briefly new experiments now under construction.

	Although these experiments span a range from 0 to greater than
$100$ {\rm GeV}/c in beam momentum, they have certain important features in
common.  In each case the source of the kaons is the interaction of
protons from a synchrotron with a fixed target.  An intense secondary
beam is created and transmitted to a decay region viewed by a
detector.  In most cases the kaons in the beam are outnumbered by
other species of particles.  These, along with the kaons which don't decay,
have to either be transmitted through insensitive regions of the
detector or somehow absorbed without doing irreparable mischief.  All
but one of the experiments use magnetic spectrometers to measure the
momenta of the charged decay products.  The neutrals are measured in
calorimeters of various types.  Scintillator hodoscopes and particle
identification devices such as atmospheric \v Cerenkov counters and
muon filters provide triggering capability.

	The experiments all exploit common $K$ decay modes for calibration
and normalization.  Very often the same modes used in this way are sources
of the backgrounds that have to be confronted.   These backgrounds
must be fought both at the trigger and analysis level.  Particle
identification, timing, geometrical, and kinematic selection reduce the
large data samples collected to manageable size.  Except in one case,
the signal events are completely reconstructed.  Typically the last
stage of the analysis is a two-dimensional plot of effective mass of
the final state particles versus a variable which reflects their
direction.  The signal is sought in the region of the $K$ mass and
small angle with respect to the beam (see e.g. Fig \ref{f: lfv}).

	All the experiments study at least one ``tune up'' process
on which to demonstrate their ability to detect rare decays.
This is a decay topologically similar to the primary object
of the experiment, but somewhat more copious.  These are often
of considerable interest in themselves.  For experiments seeking
$K_L \to \mu e$ there is $K_L \to \mu^+ \mu^-$; for $K^+ \to \pi^+
\mu^+ e^-$ there is $K^+ \to \pi^+ e^+ e^-$; for $K^+ \to \pi^+ \nu \bar \nu$
there is $K^+ \to \pi^+ \gamma \gamma$; for $K_L \to \pi^0 e^+ e^-$
there is $K_L \to \pi^0 \gamma \gamma$ or $K_L \to \gamma \gamma ee$.

\subsection{AGS-845}

	Fig.~\ref{f: 845} is a plan view of the apparatus designed by
a BNL-Yale collaboration to perform the world's first dedicated $K_L
\to \pi^0 e^+ e^-$ experiment, AGS-845.  It was optimized to detect
all-electromagnetic $K_L$ decays (the lead filter and hodoscope at the
rear were used to veto penetrating particles).  Several million $K_L$
(along with $\sim 3 \times 10^8$ neutrons) entered the 6-meter
evacuated decay region during each 1-second AGS spill.  A
single-magnet drift chamber spectrometer measured $e^{\pm}$ momenta
and a lead glass \v Cerenkov array detected $\gamma$s.  The latter
also served to measure the $e^{\pm}$ energy.  Comparing this energy
with the $e^{\pm}$ momentum distinguishes these particles from pions
and muons.  A 2-m long atmospheric hydrogen \v Cerenkov counter completed the
particle identification.

	The expected potential backgrounds to $K_L \to \pi^0 e^+ e^-$ were
$K_L \to 2 \pi^0$ or $3 \pi^0$ in which two of the $\pi^0$ undergo
Dalitz decays, and accidental coincidences of 2 $\gamma$s with $Ke3$ decays
wherein the $\pi$ is mistaken for an electron.  AGS-845 was able to
eliminate all background and set a $90\%$ c.l. limit of $B(K_L \to \pi^0
e^+ e^-) < 5.5 \times 10^{-9}$.  This represents a large improvement
in our knowledge of this process, but falls short by at least two orders of
magnitude of the Standard Model prediction for this CP-violating decay.
However, as discussed in subsection \ref{s: klp0} , in the process of setting
this
limit, AGS-845 discovered a background which may prevent this
process from ever fulfilling its potential in the study of CP-violation,
i.e. $K_L \to e^+ e^- \gamma \gamma$.

	AGS-845 also made major contributions to the study of $K_L \to
e^+ e^- \gamma$ and $K_L \to e^+ e^- e^+ e^-$.  This experiment is now
completed.  Its results are summarized in Table~\ref{t: E845sum}

\begin{table}[htb]
\centering
\caption{Results of AGS-845}
\begin{tabular}{|l|c|c|} \hline
Mode & Result &   Comments \\ \hline
$K_L \to \pi^0 e^+ e^- $ & $ < 5.5 \times 10^{-9} $ \cite{kpee845} &
search for new scalars \\
 &  & non-S.M. CP-violation \\
 &  & future: S.M. CP-violation \\ \hline
$K_L \to e^+ e^- \gamma $
& $(9.1 \pm 0.4 {+0.6 \atop -0.5}) \times 10^{-6}$  \cite{eeg845} &
 c.f. $(9.6 \pm 0.4) \times 10^{-6}$ (theory) \\
  &$\alpha_{K^*} = -0.28 \pm 0.083 {+0.054 \atop -0.034}$  &
i.e. something besides $\rho$ needed \\ \hline
$K_L \to e^+ e^- \gamma \gamma$
& $(6.6 \pm 3.2) \times 10^{-7}$ \cite{eegg845} &
c.f. $5.8 \times 10^{-7}$ for $k^* > 5 {\rm MeV}$ \\
 &  & background to $K_L \to \pi^0 e e$ \\ \hline
$K_L \to e^+ e^- e^+ e^-$
& $(3.04 \pm 1.24 \pm 0.26) \times 10^{-8}$ \cite{eeee845}
 & c.f. $ 3.6 \times 10^{-8}$\\
 &  & background to $K_L \to e^+ e^-$ \\
\hline
\end{tabular}
\label{t: E845sum}
\end{table}

\subsection{KEK-162}

	KEK-162, shown in Fig.~\ref{f: 162}, has been built by a
KEK-Kyoto collaboration to pursue $K_L \to \pi^0 e^+ e^-$ to $\sim
10^{-10}$~\cite{e162}.  It is quite similar in concept to AGS-845,
with adaptations to the lower beam energy such as a more compressed
layout, nitrogen rather than hydrogen in the \v Cerenkovs, etc.
The main innovation is the electromagnetic calorimeter
constructed of undoped CsI, a fast, bright scintillating crystal with
excellent resolution.  This calorimeter is designed to achieve
$2\%$ rms at $1$ {\rm GeV}.  Most of the light in pure CsI is emitted
with decay time
$<30$ nsec.  Fast time response is crucial since the proponents of
KEK-162 plan to expose their detector to an order of magnitude higher
$K_L$ flux than was seen by AGS-845.  The daunting rates also motivate the
design of drift chambers which feature small cells, fast gas and
custom TDCs.

	This experiment is currently setting up.

\subsection{FNAL-731/799}

	FNAL-731 was originally built to measure $\epsilon'/\epsilon$
in $K^0 \to 2 \pi$ decays.  In spite of being highly optimized for
this purpose, it has produced several estimable rare decay results.
The apparatus, built by a Chicago/Elmhurst/FNAL/Princeton/Saclay
group, is shown in Fig.~\ref{f: 731}.  Two nearly parallel $K_L$ beams
entered a 37m long evacuated decay region.  A $B_4C$ regenerator was
shuttled between the beams on a pulse to pulse basis to provide $K_S$
decays.  A plane of thin trigger scintillators was situated
approximately halfway down the vacuum decay vessel.  Following the
downstream decay region was a 4-station drift chamber dipole
spectrometer, additional trigger hodoscope planes and an 804-element lead
glass \v Cerenkov array.  Downstream of this were photon vetoes, a 3m
thick steel muon filter and, finally, muon veto hodoscopes.  An
extensive photon veto system bordered the acceptance.

	A comparison with detectors designed for lower energy
beams is instructive.  The FNAL-731 spectrometer was relatively more
compact than the typical BNL or KEK detector and so had larger
acceptance at the cost of worse charged track momentum resolution.
Conversely, the resolution for photons was better for the higher
energy experiment because of the $1/\sqrt{E}$ behavior of the stochastic
resolution term in lead glass.  In FNAL-731, the decay region was
shorter relative to the mean $K_L$ decay length, diluting to some
extent the advantage in acceptance.  One major advantage of FNAL-731
was the clean beam in which the neutrons were of the same
order as the $K_L$ instead of ten or fifty times more numerous.  This
kept the detector rates relatively low.

	A number of rare decay results were obtained from the seven-month
1987-88 run.  Subsequently, the detector was reconfigured for the
first stage of FNAL-799, a dedicated rare decay experiment.  These
changes included the removal of the regenerator, an upstream
absorber, and the trigger scintillator plane in the
vacuum decay vessel.  These resulted in a large gain in sensitivity/
incident proton.  The number of incident protons/spill was also
increased, so that the overall sensitivity/spill increased by more
than an order of magnitude.  Other modifications were
an additional muon hodoscope plane and the development of an
online processor which allowed the use of drift chamber information
in the second level trigger.  For a small portion of this run,
a pre-shower detector was installed upstream of the lead glass array
in order to improve the sensitivity to $K_L \to \pi^0 \gamma \gamma$.
The collaboration was also modified to consist of Chicago, Elmhurst,
FNAL, Illinois, Colorado, UCLA, Rutgers, and Osaka.  The experiment
ran for 10 weeks in late 1991 and early 1992.

	The results of FNAL-731/799 on rare $K$ decays thus far are
summarized in Table~\ref{t: E731sum}.  The last three (preliminary)
results come from the first stage of FNAL-799.  This run is also expected
to yield new results on $K_L \to \pi^0 e^+ e^-$, $K_L \to \pi^0 \nu
\overline \nu$, and $K_L \to \pi^0 \gamma \gamma$, as well as the first
results of this program on $K_L \to e^+ e^- \gamma$, $K_L \to \pi \pi
e e$, $K_L \to \mu \mu e e$, $K_L \to e e \gamma \gamma$, $K_L \to
\pi^0 \mu e$ and other rare decays.

\begin{table}[htb]
\centering
\caption{Results of FNAL-731/799}
\begin{tabular}{|l|c|c|} \hline
Mode & Result &  Comments \\ \hline
$K_L \to \pi^0 e^+ e^- $ & $ < 7.5 \times 10^{-9} $ \cite{pee731} &
search for new scalars, CP viol. \\ \hline
$K_S \to \pi^0 e^+ e^- $ & $ < 4.5 \times 10^{-5} $  \cite{Spee731} &
needed for interpreting $K_L\to \pi^0 ee$ \\ \hline
$K_L \to \pi^+ \pi^- \gamma $ & $  (3.19 \pm 0.16) \times 10^{-5} $
\cite{kppgfnal} & direct emission, $k^* > 20${\rm MeV}\\ \hline
$K_L \to \pi^0 \gamma \gamma $ & $(1.86 \pm 0.60 \pm 0.60) \times 10^{-6} $
 \cite{pgg731} & spectrum agrees with $\chi$PT \\ \hline
$K_L \to \pi^0 \nu \overline \nu $ & $ < 2.2 \times 10^{-4} $  \cite{pnn731} &
CP-violating \\ \hline
$K_L \to \pi^0 \mu^+ \mu^- $ & $  <1.7 \times 10^{-8} $
\cite{mdexp} & preliminary\\ \hline
$K_L \to e^+ e^- e^+ e^- $& $(4.17 \pm 0.83) \times 10^{-8} $  \cite{eeee799}
& 28 events \\ \hline
$K_L \to \mu^+ \mu^- \gamma $ & $  (3.88 \pm 0.32) \times 10^{-7} $
\cite{mdexp} & 167 events\\
\hline
\end{tabular}
\label{t: E731sum}
\end{table}

	The second stage of FNAL-799 is expected to begin taking data
in 1994 or 1995.   The experiment will be moved and a new beam line will be
built which should allow higher primary intensity.  The lead glass array
will be replaced with one consisting of undoped CsI.  The trigger planes
and photon vetoes will also be upgraded.  With these improvements, the
experiment expects to obtain sensitivities $\leq 10^{-10}$ for many rare
decays.

\subsection{NA31}

	NA31, although taking an approach to measuring $\epsilon'/\epsilon$
which is radically different from that of FNAL-731, has had similar
success in the pursuit of rare $K$ decay modes.  The detector,
shown in Fig.~\ref{f: NA31f} was built by a collaboration of
CERN, Dortmund, Edinburgh, Mainz, Orsay, Pisa, and Seigen.
The most striking difference of this detector from the others
described in this section is the absence of a magnet.  Charged
particle trajectories are determined by two planes of drift chambers,
but their energies are measured by electromagnetic and
hadronic calorimeters.  The electromagnetic calorimeter, which is
of the lead/liquid Argon type, has extremely good energy
and position resolution.   It is finely granulated and can
distinguish two photons from one if they are separated by more
than 1 cm.   This configuration results in good acceptance for many
rare decays, good energy resolution for electromagnetic
particles, but relatively poor energy resolution for pions
(e.g. the fractional resolution on a $50$ {\rm GeV}/c pion is
$\sim 9\%$).  Other disadvantages of this configuration, such
as the difficulty of distinguishing final state particles
which are close in direction, are somewhat mitigated by the
fine segmentation of the electromagnetic calorimeter.  Similarly
the lack of an energy/momentum comparison for
distinguishing electrons from heavier particles was
partially alleviated by the longitudinal segmentation of the
electromagnetic calorimeter and the by presence of the hadronic
calorimeter.  In addition there were dedicated
particle identification systems such as a transient radiation
detector and a muon filter.  Completing the apparatus were
triggering hodoscopes and photon veto counters.

	NA31 took data in 1986, 1988, and 1989.  The rare decay
results it obtained are summarized in Table~\ref{t: NA31sum}.

\begin{table}[htb]
\centering
\caption{Results of CERN NA31}
\begin{tabular}{|l|c|c|} \hline
Mode & Result &   Comments \\ \hline
$K_L \to \pi^0 e^+ e^- $ & $ < 4 \times 10^{-8} $  \cite{peeNA31} &
search for new scalars \\ \hline
$K_L \to \pi^0 \gamma \gamma $ & $(1.7 \pm 0.3) \times 10^{-6} $
 \cite{pggNA31} & spectrum agrees with $\chi$PT \\ \hline
$K_S \to \gamma \gamma $ & $  (2.4 \pm 1.2) \times 10^{-6} $ \cite{kssna} &
rate agrees with $\chi$PT \\ \hline
$K_L \to e^+ e^- \gamma $ & $ (9.2 \pm 0.5 \pm 0.5) \times 10^{-6} $
\cite{eegNA31} & cf. theory  at $(9.1-9.5) \times 10^{-6}$ \\
 & $\alpha_K = -0.28 \pm 0.13 $& i.e. something beyond $\rho$ needed \\ \hline
$K_L \to e^+ e^- e^+ e^- $ & $ (4 \pm 3) \times 10^{-8} $  \cite{dedexp} &
First observation, 2 events \\ \hline
$K^+ \to \pi^+ X^0;~X^0 \to e^+ e^-$ & $ < 6 \times 10^{-7}$ to $< 10^{-8}$
 \cite{lightNA31} &limits depend on $m_X$ and $\tau_X$ \\
\hline
\end{tabular}
\label{t: NA31sum}
\end{table}

	NA48, a successor to NA31, is now under construction.  It features
a fast liquid Krypton-based electromagnetic calorimeter and a magnetic
spectrometer for charged particle momentum measurement.  It should
have an order of magnitude better sensitivity than NA31 to most rare decays.

\subsection{AGS-777/851}

	Fig.~\ref{f: 777} shows the apparatus employed by a
BNL/PSI/Washington/Yale group in AGS Experiments 777 and 851.  The primary
objects of these were respectively a search for $K^+ \to \pi^+ \mu^+ e^- $ and
a study of $K^+ \to \pi^+ e^+ e^- $.  Other processes sought or studied
in these experiments were $\pi^0 \to \mu^+ e^-$, $\pi^0 \to e^+ e^-$,
and $A^0 \to e^+ e^-$ (where $A^0$ is a new light particle).

	A 6 {\rm GeV}/c positive beam containing about $10^7~ K^+$/AGS pulse
impinged on a 5m evacuated tank, wherein about $10\%$ of the $K^+$ decayed.
As the beam was unseparated, $K^+$ constituted only $\sim 5 \%$ of the flux,
the majority consisting of a roughly equal mixture of $\pi^+$ and protons.
The first element of the detector was a dipole run at a $p_T$ kick of $155$
{\rm MeV}/c.  This served to remove the daughter products from the hot beam
region
and to separate them according to their charge.  The beam then passed through
holes and deadened regions in subsequent detector elements.  Downstream of
the first dipole was an MWPC spectrometer with four measuring stations.
Two preceded and two followed a dipole run at $\Delta p_T~\sim 150$ {\rm MeV}/c
with sense opposite that of the first dipole.  Situated between
each pair of measuring planes was an atmospheric gas \v Cerenkov counter.
Downstream of the spectrometer were triggering hodoscopes, a
lead-scintillator sandwich electromagnetic shower detector, and an
iron/proportional tube chamber muon identifier.

	Since positive muons only were sought, the muon identifier needed
to cover only the right (+) side.  This was one of several
optimizations made possible by confining the experiment to the $\pi^+
\mu^+ e^-$ charge combination.  On the left, where electron purity was
more important than efficiency, $H_2$ was used as \v C counter gas;
whereas on on the right, where positrons had to be efficiently vetoed,
$CO_2$ was used instead.  The choice of final state was also of great
benefit to the trigger, since $e^-$ are far less common in $K^+$ decay
than are $e^+$ (the most copious source of $e^+$ in $K^+$ decay is
$K^+ \to \pi^0 e^+ \nu$ with branching ratio $0.0482$; the most
copious source of $e^-$ is $K^+ \to \pi^+ \pi^0; \pi^0 \to e^+ e^-
\gamma$ with product branching ratio $2.54 \times 10^{-3}$).

	The most dangerous backgrounds to $K^+ \to \pi^+ \mu^+ e^-$
stem from $K^+ \to \pi^+ \pi^+ \pi^-$ and $K^+ \to \pi^+ \pi^0$; $\pi^0 \to
\gamma e^+ e^-$ decays.  In the former,
this can occur through various combinations of pion decay and
misidentification.  There are also several ways in which the latter process
can mimic $K^+ \to \pi^+ \mu^+ e^-$, the worst being the case where the
$\pi^+$ is mistaken for a $\mu^+$, and the $e^+$ for a $\pi^+$.   Since
in this instance, a pion mass is misattributed to the $e^+$, the
measured 3-body effective mass can exceed $M_K{^+}$.  Thus the usefulness of
kinematic rejection is limited, so that powerful particle identification
techniques are required.

	$K^+ \to \pi^+ \pi^+ \pi^-$ and $K^+ \to \pi^+ \pi^0$; $\pi^0 \to
\gamma e^+ e^-$ decays were not totally inimical to this experiment:
they also served to calibrate and normalize it.  The former process
was used to evaluate the performance
of the particle identification systems, design the geometrical reconstruction
and kinematic fitting  procedures, etc.

	For AGS-851, the \v C gas on the right side was changed from
$CO_2$ to $H_2$.  The $K^+ \to \pi^+ \mu^+ e^-$ trigger was dropped,
and the requirement that the $\pi^+$ be detected on the right was
lifted.

	This program finished taking data in 1989.  The AGS-777 data analysis
is now complete, that of AGS-851 continues.  The results of these experiments
thus far on rare $K$ decay are summarized in Table~\ref{t: E777sum}.

\begin{table}[htb]
\centering
\caption{Results of AGS-777/851}
\begin{tabular}{|l|c|c|} \hline
Mode & Result &  Comments  \\ \hline
$K^+ \to \pi^+ e^- \mu^+ $ & $ < 2.1 \times 10^{-10} $  \cite{alee} &
$M_H > 57 ~{\rm TeV}$ \\ \hline
$K^+ \to \pi^+ e^+ e^- $ & $ (2.75 \pm 0.23 \pm 0.13) \times 10^{-7}$
\cite{alliegro} & 500 events \\
 &$\lambda  = 0.105 \pm 0.035 \pm 0.015$  &
suggests $K_S \to \pi^0 e^+ e^- $ small\\ \hline
$K^+ \to \pi^+ X^0;~X^0 \to e^+ e^-$ & $ < 1.1 \times 10^{-8}$
\cite{alliegro} & $150 < m_{ee} < 340 {\rm MeV}$ \\
     & $ < 4.5 \times 10^{-7}$
\cite{H777} & $100 {\rm MeV} < m_{ee} $ \\
 &  & for $ \tau_{X} < 10^{-13} $ sec. \\
\hline
\end{tabular}
\label{t: E777sum}
\end{table}

	Subsequent to the completion of this program, the AGS Booster
came on line, and the prospect of much greater available $K^+$ flux
motivated the proposal of a successor experiment, AGS-865.  The
institutions collaborating are BNL, INR-Moscow, Dubna, New Mexico,
PSI, Basel, Pittsburgh, Tbilisi, Yale, and Zurich.  Since AGS-777
was limited primarily by beam-associated background, a new beam has
been designed to yield seven times more $K^+$ with no greater
random rates than those of its predecessor.  The detector, shown in
Fig.~\ref{f: 865}, is very similar to that of AGS-777.  The geometrical
acceptance has been increased by about a factor three with respect to
that of the earlier detector, however, and the muon identifier covers
both sides of the apparatus.  Improvements to the background rejection
power of the experiment include a fourth PWC plane at each measuring
station, the use of aluminum HV wires to reduce multiple scattering,
an upgraded electromagnetic calorimeter, finer longitudinal sampling
in the muon identifier, etc.

	The increases in the $K^+$ flux, the geometric acceptance, the
triggering and reconstruction efficiencies and in running time are
expected to yield a factor 70 improvement in the sensitivity of
AGS-865 over that of AGS-777/851.  This would allow a $K^+ \to \pi^+
\mu^+ e$ sensitivity of $1.3 \times 10^{-12}$/event for example.
At the same time samples of tens of thousands of decays such as
$K^+ \to \pi^+ e^+ e^-$, $ \pi^+ \mu^+ \mu^-$ and $\pi^+ \gamma
\gamma$ should be accumulated.  There are also a number of other
interesting processes which could be studied with special runs and/or modest
upgrades to the detector.  These include CP-violating asymmetries in
$K^{\pm} \to \pi^{\pm} \pi^+ \pi^-$, T-violating polarization in
$K^+_{\mu 3}$ decay, and parity-violating $\mu^+$ polarization asymmetry
in $K^+ \to \pi^+ \mu^+ \mu^-$.

\subsection{KEK-137}

	KEK-137, a Tohoku/Tokyo/Kyoto/KEK collaboration, is one of the
two most recent $K_L \to$ 2-lepton experiments.  These experiments
have had to meet very significant challenges to achieve sensitivities
significantly better than $10^{-10}$/event.  The fraction of $K_L$
that decay and can be accepted in a practical sized apparatus is
typically $\sim 1-2 \times 10^{-3}$.  To get enough $K_L$, the
experiments have to work in the forward or near-forward direction
where the neutron flux is $10 - 100$ times higher than the $K_L$.
This translates into neutral beam fluxes of order $10^9$ per spill.
Chamber and trigger plane rates are typically many MHz, and yet they
must perform extremely well because to reject background,
searches for $K_L \to \mu e$ must have excellent kinematic
resolution.   Since the primary background, $K_{e3}$ decay followed by $\pi
\to \mu$ via decay or misidentification, occurs at the few $\%$ level
and is topologically identical to the signal, particle identification
power is of limited value.  Most of the background rejection comes from
kinematic and geometrical cuts.  One exploits the fact that in the
absence of measuring errors, the $e$ ``$\mu$'' pairs have effective
masses less than $M_K -8.4$ {\rm MeV}.  Both recent experiments
kinematically and geometrically over-constrain their events.  The
momentum of each track is measured twice spectrometrically and in the
case of muon candidates, once more via range.

	Here again, the background decays ($K_{e3}$ and for the
$K_L \to \mu^+ \mu^-$ case, $K_{\mu 3}$) are not all bad.
In this case they serve to calibrate the particle identification
devices.  The relatively copious $K_L \to \pi^+ \pi^-$ are used
to normalize the experiments as well as to calibrate
the spectrometers.

	The double-arm spectrometer used by KEK-137 to search
for $K_L \to \mu e$, $K_L \to e^+ e^-$, $K_L \to \mu^+ \mu^-$, and
other rare decay modes is shown in Fig.~\ref{f: 137}.  A beam of $\sim
10^7~K_L$/pulse was made by directing a $1-2 \times 10^{12}$ proton
beam from the KEK PS onto a 12 cm-long Cu target.  The beam passed
through a number of collimators and sweeping magnets into a 10m-long
evacuated decay volume.  About $8\%$ of the $K_L$ between $2$ and
$8$ {\rm GeV} decayed in this volume.  The rest of the neutral beam,
which included about $10^9$ neutrons and $\gamma$s per spill, was
conducted in vacuum between the two spectrometer arms.   The
$p_T$ kick on each arm was $238$ {\rm MeV}/c, divided equally between
the two dipoles.
Daughter tracks in the Jacobean peak of the desired two-body
reactions were consequently bent approximately parallel to
the arm axes.  Imposing a parallelism requirement greatly reduced
the relative number of three-body decays accepted by the trigger.
The resolution of this spectrometer for the calibration $K_L \to
\pi^+ \pi^-$ decays was $1.3$ {\rm MeV}/$c^2$.  As mentioned above,
there were separate momentum measurements by the front and rear
sections of the spectrometer and a third, coarse, measurement via
a muon range array.  The multiple measurements were primarily aimed
at rejecting pions which decayed to muons in the detector.  These
were more dangerous than punch-through pions since the decay could
disturb the momentum measurement as well as the particle
identification.  Electrons were identified via atmospheric
\v Cerenkovs filled with atmosphere, and with planes
of lead-scintillator shower counters.

	This experiment ran for a period of about $2 {1 \over{2}}$ years,
finishing data-taking in May, 1990.  The results are summarized in
Table~\ref{t: E137sum}.

\begin{table}[htb]
\centering
\caption{Results of KEK-137}
\begin{tabular}{|l|c|c|} \hline
Mode & Result &  Comments  \\ \hline
$K_L \to \mu e$ & $ 9.7 \times 10^{-11} $  \cite{mue137} &  \\ \hline
$K_L \to e^+ e^-$ & $ 1.6 \times 10^{-10} $ \cite{mue137}  &
 one event in signal region\\ \hline
$K_L \to \mu^+  \mu^-$ & $ (7.9 \pm 0.6 \pm 0.3) \times 10^{-9} $
\cite{mumu137} & $178$ events \\ \hline
$K_L \to e^+ e^- e^+ e^-$ & $ (6 \pm 2 \pm 1) \times 10^{-8} $
\cite{eeee137} & partial
reconstruction \\
\hline
\end{tabular}
\label{t: E137sum}
\end{table}

\subsection{AGS-791}

	The second of the two-lepton experiments was AGS-791, a
collaboration of UC-Irvine, UCLA, LANL, U of Pennsylvania, Stanford,
Temple, and William \& Mary.  The detector, shown in Fig.~\ref{f: 791},
had many similarities to that of KEK-137: a double arm - double measuring
spectrometer, electron ID via atmospheric \v C counter plus electromagnetic
shower counter, muon range array, etc.  However there were significant
differences.  The primary beam was about twice that of the KEK accelerator,
and the daughter tracks entering the apparatus were roughly twice as
stiff as those of KEK-137.  Thus the muon identifier was considerably
thicker and the \v Cerenkov needed to be filled with a gas of lower index
of refraction (He-Ne mixture vs air).  In E791, the arms shared
their spectrometer magnets, and these were set to opposite polarities
(the $p_T$ kick of each was $\sim 300$ {\rm MeV}/c).   Other differences from
KEK-137
included a shorter decay volume but larger geometrical acceptance,
lead glass instead of lead-scintillator sandwich counters, finer sampling
in the muon identifier, etc.  The $K_L$ flux impinging on the
detector ($\sim 5 \times 10^7$/ 1-second spill) was the highest yet
used in a $K_L$ decay experiment, and put severe requirements on
the triggering and data acquisition systems.

	The analysis flow was also similar to that of their KEK
competitors.  Normalization and spectrometer calibration was done
with $K_L \to \pi^+ \pi^-$ decays, calibration of particle identification
devices via $K_{\ell 3}$ decays.

	In the course of analyzing the experiment, an unanticipated
potential background to $K_L \to \mu e$ was discovered.  This was
$K_L \to \pi^{\pm} e^{\mp} \nu$ decay in which both charged
daughters were misidentified ($\pi$ as $e$, $e$ as $\mu$).  Since
a much higher mass is attributed to the electron, the reconstructed
two-body effective mass can span the region of the signal.  Fortunately
the particle identification power of the experiment was sufficient
to reduce this background to $\leq 10^{-12}$.

	The data was taken in three runs over the period 1988-1990.
The sensitivities reached were the highest ever attained in a $K$
decay experiment.  The results are summarized in Table~\ref{t: E791sum}.

\begin{table}[htb]
\centering
\caption{Results of E791}
\begin{tabular}{|l|c|c|} \hline
Mode & Result &   Comments  \\ \hline
$K_L \to \mu e$ & $ 3.3 \times 10^{-11} $ \cite{mue791} & Most sensitive
$K$ experiment yet \\ \hline
$K_L \to e^+ e^-$ & $ 4.7 \times 10^{-11} $  \cite{mue791} & \\ \hline
$K_L \to \mu^+  \mu^-$ & $ (7.0 \pm 0.5) \times 10^{-9} $  \cite{mumu791} &
$718$ events \\
\hline
\end{tabular}
\label{t: E791sum}
\end{table}

	After the completion of data taking on AGS-791, a
collaboration of UC-Irvine, Stanford, Temple, Texas, and William \&
Mary proposed a successor experiment, AGS-871.  A schematic of their
detector is show in Fig.~\ref{f: 871}.  Although many of the elements
of the previous experiment will be reused, there are important
differences in the design.  The most striking of these is that instead
of allowing the beam to pass unimpeded between the arms of the
detector as in AGS-791, here it is stopped by a plug in the first
spectrometer magnet.  This allows larger geometric acceptance and
lower rates in the downstream chambers and particle identification
devices, at the expense of some increase in rates in the chambers near
the plug.  To help cope with this, the forward drift chambers will be
replaced by high-rate straw trackers.  A second important change is an
adjustment in the spectrometer magnet fields so that there is a net
$\Delta p_T$ of $220${\rm MeV}/c, to bend the two-body decay daughter tracks
parallel to the beam, allowing a faster, more effective first level
trigger and simpler \v Cerenkov optics.  The over-bend ($\Delta p_T
= 440$ - $220${\rm MeV}/c) also leads to about a $20\%$
improvement in the two-body effective mass resolution.  Other changes
include a longer decay volume, increased chamber redundancy, the use
of $H_2$ in the \v Cerenkov, better muon range resolution, and
upgraded triggering and data acquisition systems.  The $K_L$ beam line
will be lengthened to allow improved collimation.

	Assuming a four-fold increase in the available
AGS intensity, the overall improvement in sensitivity expected is
about a factor $20$.  This implies single event sensitivities
better than $10^{-12}$.  Thus a sample of over $10,000$ $K_L \to \mu^+
\mu^-$ will be accumulated as well as a few examples
of $K_L \to e^+ e^-$.

\subsection{AGS-787}

	The apparatus built by a BNL/Princeton/TRIUMF collaboration
to carry out the first stage of AGS-787 is shown in Fig.~\ref{f: 787}.
The solenoidal configuration, unique among the detectors described in
this section, was mandated by the problematic signature of $K^+ \to
\pi^+ \nu \overline \nu$.  A $\pi^+$ is hardly a novelty in the final state
of a $K^+$ decay and since it alone of the three daughters is
detectable, one has only the weak kinematic constraint $p_{\pi} \leq
(M_K^2 - M_{\pi}^2)/2M_K$.  What makes the experiment possible is the
good signature of the {\sl backgrounds}.  The leading
$K^+$ decays by far are $K_{\pi2}$ and $K_{\mu2}$, each of which features a
single charged track of unique cm momentum ($205$
{\rm MeV}/c and $236$ {\rm MeV}/c
respectively).  Thus with good kinematic resolution one can
reject the two-body backgrounds by a large factor.
In addition one can veto on the photons from $K_{\pi2}$
and on the identity of the muon in $K_{\mu2}$.  Other potential sources of
background (e.g. $K^+ \to \mu^+ \nu \gamma$, $K_{\mu3}$, $K_{\pi3}$, etc.)
are less copious and all have combinations of at least two of the
three `handles' mentioned above: kinematic signature, detectable extra
tracks, and charged daughter $\neq \pi^+$.

	Like all previous searches for $K^+ \to \pi^+ \nu \overline \nu$,
AGS-787 employs a stopping $K^+$ beam.  This allows direct
access to the kinematic features of signal and background.  Other advantages
are the feasibility of large geometric acceptance and veto
hermiticity, the powerful particle identification techniques
available at low energy, and the very good ratio of useful $K^+$
decays to unwanted beam particles.  The latter results from the
pure separated beams which are practical at low energies and
the fact that one can stop a relatively large fraction of a low
energy $K^+$ beam.  This is quite important in an experiment in which
the signature is a single unaccompanied $\pi^+$.

	In the design of AGS-787, great efforts were made to minimize
the presence of ``dead'' material.  Energy deposited or interactions
undergone in such material can compromise the veto or confound the
particle identification.  For example, a $\pi^+$ whose scatter in the
stopping target goes undetected can defeat the kinematic rejection of
$K_{\pi2}$, if the $\pi^0$ decay photons are also missed.

	The background due to such ``down-shifted'' $K_{\pi2}$ decays led
the experimenters initially to concentrate on the kinematic region with
$p_{\pi^+} > 205$ {\rm MeV}/c where the only sources of $\pi^+$ more copious
than the signal are are $K^+ \to \pi^+ e^+ e^-$ and $K^+ \to \pi^+
\gamma \gamma$.  These have branching ratios of $\sim 3 \times 10^{-7} $
and $<10^{-6}$ respectively.  The rejection of electromagnetic
particles is extremely good in AGS-787 (e.g. only $1$ or $2$ of $10^6~
\pi^0$ are missed), so that these do not constitute a significant
problem.  In fact in this region, the most difficult backgrounds
have proved to be $K^+ \to \mu^+ \nu$ with muons which interact
or enter the dead regions of the detector before making their range.

	Thus far, AGS-787 has operated at $K^+$ stopping rates up to $\sim
300,000$/ beam spill.  To obtain this rate, about $1.5M$ $800$ {\rm MeV}/c
$K^+$ from the LESB1 were directed onto a BeO degrader 54cm in length.  The
$K^+$ were accompanied by an approximately equal number of protons and
by about $3M$ $\pi^+$.  The $K^+$ emerged from the degrader with about
$100$ {\rm MeV} of energy and came to rest in a highly segmented
scintillating fiber target~\cite{app787}.  $K^+$ were required to decay
at least $2$ nsec after stopping to be accepted.  Charged daughters
which leave the target within about $30^0$ of the plane transverse to
the beam were tracked a cylindrical drift chamber onto which a field
of 1T was imposed.  These then entered a cylindrical array
of plastic scintillation counters and PWCs (range stack) which served
to measure their residual energy and range.  The range stack was read out
on both upstream and downstream ends so that the detected particles could
be localized in three dimensions.  The counters were instrumented with
500 MHz transient recorders which recorded all scintillation light over
an interval of about $10~\mu sec$.  This allowed the characteristic
$\pi \to \mu \to e$ sequence to be observed in the stopping counter.
This is a powerful signature for $\pi^+$, which was defeated in $<1$
out of $10^5$ cases.  Comparisons among range, energy, and momentum of the
daughter track also constitute a very effective particle identification
technique.   Surrounding the range stack was a cylindrical array of
lead-scintillator sandwich counters (barrel veto) which veto
gammas in the central region.  The gamma veto was completed by
lead-scintillator sandwich end-cap counters.

	Calibration and normalization of the experiment was done primarily
via $K_{\pi2}$ and $K_{\mu2}$.

\begin{table}[htb]
\centering
\caption{Results of E787}
\begin{tabular}{|l|c|c|} \hline
Mode & Result &   Comments  \\ \hline
$K^+ \to \pi^+ X^0 $ & $< 1.7 \times 10^{-9} $ \cite{pnn787} &
familon, etc. search; \\ \hline
$K^+ \to \pi^+ \nu \overline \nu $ & $< 5.2 \times 10^{-9} $
\cite{pnn787,pnn2} &
constrains new physics \\ \hline
$K^+ \to \pi^+ \mu^+ \mu^- $ & $ <2.3 \times 10^{-7}$ \cite{pmm787} &
should be discovered in '89-'91 data\\ \hline
$K^+ \to \mu^+ \nu \mu^+ \mu^-. $ & $ <4.1 \times 10^{-7}$ \cite{pmm787} &
Higgs hunting ground \\ \hline
$K^+ \to \pi^+ \gamma \gamma $ & $ < 10^{-6}$ \cite{pgg787} &
should be discovered in '89-'91 data\\ \hline
$K^+ \to \pi^+ X^0;~X^0 \to \gamma \gamma $ & $ \leq 10^{-7}$ \cite{pgg787} &
$m_{X^0} \leq 90$ {\rm MeV}/$c^2$ \\
\hline
\end{tabular}
\label{t: E787sum}
\end{table}

	The results of AGS-787 are summarized in Table~\ref{t: E787sum}.
These results are based on data collected in an 1988 engineering run and
in the first major physics run (1989).   There were runs of about equal
sensitivity
in 1990 and 1991.  During the 1989 run it was found that the
instrumentation of the experiment was sufficient to reject background
in the kinematic region below the $K_{\pi 2}$ as well as above it.
Thus far the acceptance in the two kinematic regions has proved to
be about equal.

	In 1989 a major upgrade of the beam and detector was approved
with the aim of reaching the $10^{-10}$ sensitivity level necessary
to probe the Standard Model predictions for $K^+ \to \pi^+ \nu \overline \nu$.
The collaboration was augmented by groups from INS/Tokyo and KEK.
A new low energy separated beam (LESB3) which provides much improved $K^+$
flux and purity was constructed.  This beam was designed to exploit the AGS
upgrade to deliver up to $1.5 \times 10^7~800${\rm MeV}/c $K^+$ with a
$K^+/\pi^+$
ratio of $2/1$.  Extensive detector improvements were also
undertaken.  These include upgrades to the $K^+$ and $\pi^+$ measuring
devices, the photon vetoes, the electronics and data acquisition
system.  A new, brighter, stopping target was constructed out of
close-packed 5mm square cross-section scintillating fibers.  A
new central drift chamber with only ${1 \over 5}$ the mass of its
predecessor is being constructed.  In the range stack, the scintillator
read out granularity is being increased and the embedded proportional
chambers are being replaced by far less massive straw chambers.
Pure CsI end cap vetoes are being built to replace the current
lead-scintillator sandwich devices.   A pure CsI liner will be
installed inside the current barrel veto.  A number of supplementary
vetoes will increase the hermiticity of the the detector.  Transient
recorders will be installed on the target and on all vetoes.  Finally,
a new trigger and data acquisition system, capable of taking ten times
the previous rate is being developed.

%\clearpage

\section{Conclusions}

The study of rare kaon decays continues to be a crucial arena for the
testing of electroweak theory. The decays $K \ra \pi \nu \overline{\nu}$,
can be computed reliably, and their study will
yield valuable information on the CKM parameters $\rho$ and $\eta$, that
will eventually permit us to test the three generation structure of the
standard model.  Other rare decays that are sensitive to these parameters
are $K_L \ra \pi^0 \ell^+ \ell^-$, $K^+ \ra \pi^+ \ell^+ \ell^-$ and perhaps
$K_L \ra \mu^+ \mu^-$ as well. Of course, these modes also constitute
ideal candidates to search for new physics in the form of deviations from
the standard model expectations. This is particularly true for CP
violation.

The search for the forbidden lepton flavor violating decays constitutes
one of the simplest and most cost-effective ways to constrain interactions
beyond the minimal standard model. The present level of sensitivity of
these experiments is already testing energy scales that we will not be able
to probe directly for many years.

In the process of searching for the very rare and forbidden decay modes of
the $K_L$ and the $K^\pm$, large samples of other, less rare, decays are
being collected. Among them are the radiative decays that are dominated by
long distance contributions. Their detailed understanding is crucial in
the effort to use modes like $K_L \ra \pi^0 e^+ e^-$ or $K_L \ra \mu^+ \mu^-$
to measure the parameters $\rho$ and $\eta$. These radiative decay modes
are also interesting in their own right, in that they allow us to  test
the framework of chiral perturbation theory, and in that context they
provide information on the non-perturbative aspects of the strong
interactions.

There are a number of accelerator developments under way to augment
the supply of kaons for these studies.
The AGS upgrade is well along.  It will provide a fourfold increase
in what is already the world's most intense source of kaons.  Construction
has begun on the Fermilab Main Injector which promises to provide a
source of comparable intensity at higher energy.  Construction of the
DA$\Phi$NE $\phi$ storage ring at Frascati is also under way.  This
facility will provide a somewhat less intense source of kaons, but
one in which the initial state of the kaons can be tightly controlled.
In the somewhat more distant future, facilities such as TRIUMF
Laboratory's proposed KAON complex can provide further large increments
in sensitivity.

The field of rare kaon decays has provided many discoveries of
the highest importance.  From an historical perspective, the
reach in sensitivity of the present and near-future experiments is
quite large.  It would be very surprising if further discoveries
did not await these initiatives.

{\bf Acknowledgements}

We wish to acknowledge useful conversations and other assistance from
W.~Bardeen, A.~Barker, L.~Bergstrom, A.~Buras, H.~Y.~Cheng,
S.~Dawson, J.~F.~Donoghue, C.~Q.~Geng, E.~Golowich,
T.~Han, X.~G.~He, J.~Kambor, S.~Kettell, A.~El-Khadra,
H.~Ma, W.~Marciano, W.~Morse, K.~Ohl, J.~Prades,
C.~Quigg, E.~Ramberg, S.~Willenbrock,
B.~Winstein, T.~Yamanaka, and M.~Zeller.

This work was supported by the U.S. Department of Energy under
contracts DE-AC02-76CH00016 and DE-AC02-76CH03000.

\newpage

\listoftables

\newpage

\begin{figcap}

\item{Results from recent searches for lepton flavor violation in K decay.
a) $m(\mu e)$ vs $\theta_{lineup}^2$ for $K_L \ra \mu e$ candidates from
KEK-137;  b) $m(\mu e)$ vs $p_T^2$ for $K_L \ra \mu e$ candidates from
AGS-791; c) $m(\pi \mu e)$ vs vertex quality variable for
$K^+ \ra \pi^+ \mu^+ e^-$ candidates from AGS-777.}
\label{f: lfv}

\item{Box diagrams giving rise to $K_L \ra \mu^\pm e^\mp$ and
$K \ra \pi \mu^\pm e^\mp$ in models with massive neutrinos.}
\label{f: bnn}

\item{Short distance contributions to $K \ra \pi \nu \overline{\nu}$.
The full circle represents the effective one-loop $sdZ$ coupling.}
\label{f: sdd}

\item{Potential long distance contributions to
$K^+ \ra \pi^+ \nu \overline{\nu}$.}
\label{f: impd}

\item{Search regions for $K^+ \ra \pi^+ \nu \bar \nu$ from AGS 787.
a) $p_{\pi^+} > p_{K \pi 2}$ and b) $p_{\pi^+} < p_{K \pi 2}$.}
\label{f: 787dat}

\item{Loop diagrams  that give rise to $K_L \ra \pi^0 \gamma \gamma$ at
order ${\cal O}(p^4)$. The same diagrams without the $\pi^0$ line give
rise to $K_S \ra \gamma \gamma$. The {\bf X} represents a vertex from
Eq.~\ref{wlt}.}
\label{f: kgg}

\item{Examples of ${\cal O}(p^6)$ contributions to
$K_L \rightarrow \pi^0 \gamma \gamma$ in VMD models.
a) Pole diagrams and b) Direct weak counter-terms.
In both cases the {\bf X} represents a weak transition, but in (b) it is
${\cal O}(p^6)$.}
\label{f: poles}

\item{Rate for $K_L \ra \pi^0 \gamma \gamma$. a) Theoretical
spectra.  The solid line is the ${\cal O}(p^4)$ $\chi$PT ($\alpha_V = 0$)
result, and the dashed and dotted lines show the range for the values of
$\alpha_V$
given in the text.  Phase space and pure VMD spectra are also shown; b)
Data from Ref.~\cite{pggNA31} (solid histogram), compared with
${\cal O}(p^4)$ result (dotted histogram).  The latter has been
multiplied by the experimental acceptance (shown as crosses).  Dashed
histogram is calculated background.}
\label{f: rkpgg}

\item{Pole diagrams that give rise to $C(z,\nu)$ in $K^\pm \ra \pi^\pm
\gamma \gamma$.}
\label{f: kpgep}

\item{Pole diagrams contributing to $K_L \ra \gamma \gamma$}
\label{f: akgg}

\item{Pole diagrams contributing to the direct emission in
$K_L \ra \pi^+ \pi^- \gamma$.}
\label{f: mkppg}

\item{Short distance diagrams giving rise to $K_L \ra \ell^+ \ell^-$
and $K^+ \ra \pi^+ \ell^+ \ell^-$. The full circle represents the effective
one-loop $sdZ$ vertex.}
\label{f: sdll}

\item{Dominant long distance contribution to $K_L \ra \ell^+ \ell^-$.
The vertical dashed line represents the cut to obtain the absorptive part.}
\label{f: ldll}

\item{Diagrams contributing to $K \ra \pi \gamma^*$ at ${\cal O}(p^4)$.
The full circle represents a vertex from Eq.~\ref{slf}, whereas the full
box represents a vertex from Eq.~\ref{wlfn}. The {\bf X} is a ${\cal O}(p^2)$
weak transition from Eq.~\ref{wlt}.}
\label{f: kpgee}

\item{Calculated decay distribution for $K^+ \ra \pi^+ e^+ e^-$,
for values of $w_+$ mentioned in the text.}
\label{f: kpllr}

\item{Fit of $K^+ \ra \pi^+ e^+ e^-$ spectrum and branching ratio
to predictions of $\chi$PT.  From Ref.~\cite{alliegro}.}
\label{f: zeller}

\item{Short distance contributions to the direct CP violation in
$K_L \ra \pi^0 e^+ e^-$. Again, the full circle represents effective
one-loop couplings.}
\label{f: skee}

\item{Two photon contribution to the CP conserving absorptive part of
$K_L \ra \pi^0 e^+ e^-$.}
\label{f: ldee}

\item{Background to the process $K_L \ra \pi^0 e^+ e^-$.}
\label{f: bkee}

\item{Plan view of AGS-845 detector}
\label{f: 845}

\item{Schematic of KEK-162 detector}
\label{f: 162}

\item{Elevation view of FNAL-731 detector}
\label{f: 731}

\item{Schematic layout of NA31 detector}
\label{f: NA31f}

\item{Plan view of AGS-777 detector}
\label{f: 777}

\item{Schematic of proposed AGS-865 detector}
\label{f: 865}

\item{Plan view of KEK-137 detector}
\label{f: 137}

\item{Plan view of AGS-791 detector}
\label{f: 791}

\item{Schematic of proposed AGS-871 detector}
\label{f: 871}

\item{Side elevation view of AGS-787 detector}
\label{f: 787}

\end{figcap}


\clearpage

\begin{thebibliography}{999}


%{Introduction}

\bibitem{bryman}{D.~Bryman, {\it Int. Jour. of Mod. Phys.} {\bf A4} 79 (1989).}

\bibitem{rhl}{J.~Hagelin and L.~Littenberg, {\it Prog. Part. Nucl. Phys.}
{\bf 23} 1 (1989).}

\bibitem{rbcfp}{R.~Battiston, {\it et. al.},
{\it Phys. Rep.} {\bf 214} 293 (1992).}

\bibitem{rrw}{J.~Ritchie and S.~Wojcicki, {\it Rev. of Mod. Phys.}
to appear.}

\bibitem{rdhv}{J.~Donoghue, B.~Holstein and G.~Valencia, {\it Int. Jour.
of Mod. Phys.} {\bf A2} 319 (1987).}

\bibitem{ekre}{U.~Meissner, BUTP-93/01 to appear in {\it J. Phys. G:
Nucl. Part. Phys.}}

\bibitem{mar}{W.~Marciano, Rare Decay Symposium, Ed. D.~Bryman {\it et. al.},
World Scientific 1 (1988).}

\bibitem{wolf}{B.~Winstein and L.~Wolfenstein, {\it Rev. of Mod. Phys.}
to appear.}

\bibitem{buras}{A.~Buras and M.~Harlander, {\it Review Volume on Heavy
Flavors}, ed. A.~Buras and M.~Lindner, World Scientific, Singapore (1992).}

%{Kaon Decays in the Standard Model}

\bibitem{inami}{T.~Inami and C.~S.~Lim, {\it Prog. Theo. Phys.}
{\bf 65} 297 (1981); E.{\bf 65} 1772 (1981).}

\bibitem{burasb}{G.~Buchalla, A.~Buras and M.~Harlander, {\it Nucl. Phys.}
{\bf B349} 1 (1991).}

\bibitem{wolkm}{L.~Wolfenstein, {\it Phys. Rev. Lett.} {\bf 51} 1945 (1983).}

\bibitem{vusex}{H.~Leutwyler and M.~Roos, {\it Z. Phys.} {\bf C25} 91 (1984).}

\bibitem{vcbex}{M. Neubert, {\it Phys. Lett.} {\bf B264} 455 (1991);
G. Burdman, {\it Phys. Lett.} {\bf B284} 133 (1992).}

% reference for $b \to u/b \to c$.

\bibitem{buobc}{D.~Cassel,talk at DPF meeting FNAL Nov. 1992.}

%{Chiral Perturbation Theory}

\bibitem{cpta}{S.~Weinberg, {\it Physica} {\bf 96A} 327 (1979).}

\bibitem{cptb}{J. Gasser and H. Leutwyler, {\it Ann. Phys.}
{\bf 158} 142 (1984); J. Gasser and H. Leutwyler, {\it Nucl. Phys.}
{\bf B250} 465  (1985).}

\bibitem{cptc}{H.~Georgi, {\it Weak Interactions and Modern Particle Physics}
Benjamin, New York, (1984).}

\bibitem{dghb}{J.~Donoghue, E.~Golowich and B.~Holstein, {\it Dynamics of the
Standard Model} Univ. Press, Cambridge UK (1992).}

\bibitem{rsld}{J.~Bijnens, G.~Ecker and J.~Gasser, CERN-TH-6625/92
and references therein.}

\bibitem{holfpi}{B.~Holstein, {\it Phys. Lett.} {\bf 244B} 83 (1990).}

\bibitem{wzw}{J.~Wess and B.~Zumino, {\it Phys. Lett.} {\bf 37B} 95 (1971);
E.~Witten, {\it Nucl. Phys.} {\bf B223} 422 (1983).}

\bibitem{cron}{J.~A.~Cronin, {\it Phys. Rev.} {\bf 161} 1483 (1967).}

\bibitem{dghpr}{J.~Donoghue, E.~Golowich and B.~Holstein, {\it Phys. Rep.}
{\bf 131} 319 (1986).}

\bibitem{devlin}{T.~J.~Devlin and J.~O.~Dickey, {\it Rev. of Mod. Phys.}
{\bf 51} 237 (1979).}

\bibitem{kambor}{J.~Kambor, J.~Missimer and D.~Wyler, {\it Nucl. Phys.}
{\bf B346} 17 (1990).}

\bibitem{rafao}{G.~Ecker, A.~Pich and E.~de~Rafael, {\it Nucl. Phys.}
{\bf B291} 692 (1987).}

\bibitem{eprt}{G.~Ecker, A.~Pich and E.~de~Rafael, {\it Nucl. Phys.}
{\bf B303} 665 (1988).}

\bibitem{chengan}{H.~Y.~Cheng, {\it Phys. Rev.} {\bf D42} 72 (1990).}

\bibitem{bepan}{J.~Bijnens, G.~Ecker and A.~Pich, {\it Phys. Lett.}
{\bf 286B} 341 (1992).}

\bibitem{bbg}{W.~Bardeen, A.~Buras and J.~Gerard, {\it Phys. Lett.}
{\bf 180B} 133 (1986); {\bf 211B} (1988) 343; {\it Nucl. Phys.}
{\bf B293} 787 (1987).}

\bibitem{goctet}{A.~Pich and E.~de~Rafael, {\it Nucl. Phys.} {\bf B358}
311 (1991).}

\bibitem{vmdl}{J.~Donoghue, C.~Ramirez and G.~Valencia, {\it Phys. Rev.}
{\bf D39} 1947 (1989); G.~Ecker, {\it et. al.},
{\it Nucl. Phys.} {\bf B321} 311 (1989).}

\bibitem{qml}{J.~Bijnens, C.~Bruno and E.~de~Rafael, CERN-TH-6521/92,
and references therein.}

\bibitem{donho}{J.~Donoghue and B.~Holstein, {\it Phys. Rev.} {\bf D40}
2378 (1989); {\it ibid.} 3700; and references therein.}

\bibitem{wdmw}{G.~Ecker, A.~Pich and E.~de~Rafael, {\it Phys. Lett.}
{\bf 237B} 481 (1990).}

\bibitem{hyc}{H.~Y.~Cheng, {\it Phys. Rev.} {\bf D42} 3850 (1990);
{\it Phys. Lett.} {\bf 238B} 399 (1990).}

\bibitem{bruno}{C.~Bruno and J.~Prades, CPT-92/P 2795 (1992).}

%\bibitem{wdres}{G.~Ecker, J.~Kambor and D.~Wyler, CERN-TH-6610/92.}

%{Lepton family number violating decays}

\bibitem{buwy}{W.~Buchm\"{u}ller and D.~Wyler, {\it Nucl. Phys.} {\bf B268}
621 (1986).}

\bibitem{pdb}{Particle Data Group 1992, {\it Phys. Rev.} {\bf D45} No.11
part II.}

\bibitem{mue137}{T.~Akagi {\it et. al.}, {\it Phys. Rev. Lett.} {\bf 67}
2614 (1991).}

\bibitem{mue791}{K.~Arisaka {\it et. al.}, {\it Phys. Rev. Lett.}
{\bf 70} 1049 (1993).}

\bibitem{cahara}{R.~Cahn and H.~Harari, {\it Nucl. Phys.} {\bf B176}
135 (1980).}

\bibitem{pati}{J.~Pati and H.~Stemnitzer, {\it Phys. Lett.} {\bf 172B}
441 (1986).}

\bibitem{shanker}{O.~Shanker, {\it Nucl. Phys.} {\bf B206} 253 (1982).}

\bibitem{mueexp}{R.~Bolton, {\it et. al.}, {\it Phys. Rev.} {\bf D38}
2121 (1988).}

\bibitem{acpak}{A.~Acker and S.~Pakvasa, {\it Mod. Phys. Lett.} {\bf A7}
1219 (1992).}

\bibitem{marhn}{W.~Marciano and A.~Sanda, {\it Phys. Rev. Lett.}
{\bf 38} 1512 (1977); W.~Marciano, {\it Phys. Rev.} {\bf D45}
R721 (1992).}

\bibitem{babrbe}{A.~Barroso, G.~Branco and M.~Bento, {\it Phys. Lett.}
{\bf 134B} 123 (1984).}

\bibitem{lasasc}{P.~Langacker, S.~Uma~Sankar and K.~Schilcher, {\it Phys. Rev.}
{\bf D38} 2841 (1988).}

\bibitem{dava}{S.~Dawson and G.~Valencia, Fermilab-Pub-93/024-T.}

\bibitem{alee}{A.~M~Lee, {\it et. al.}, {\it Phys. Rev. Lett.} {\bf 64}
165 (1990).}

\bibitem{diam}{A.~Diamant-Berger, {\it et. al.}, {\it Phys. Lett.}
{\bf 62B} 485 (1976).}

\bibitem{beyondsm}{SUSY: B.~Campbell, {\it Phys. Rev.} {\bf D28} 209 (1983);
Technicolor: E.~Eichten, {\it et. al.}, {\it Phys. Rev} {\bf D34} 1547
(1986); Leptoquarks: L.~Hall and L.~Randall, {\it Nucl. Phys.} {\bf B274}
157 (1986); Horizontal interactions: W.~S.~Hou and A.~Soni, {\it Phys. Rev.}
{\bf D35} 2776 (1987); Multiscalar models: H.~Haber and Y.~Nir, {\it Nucl.
Phys.} {\bf B335} 363 (1990); G.~B\'{e}langer, C.~Q.~Geng and P.~Turcotte,
{\it Phys. Rev.} {\bf D46} 2950 (1992); A.~Antaramian, L.~Hall and A.~Rasin,
{\it Phys. Rev. Lett.} {\bf 69} 1871 (1992). A recent review with more
references is P.~Langacker in 1990 Snowmass proceedings.}

%{$K_L \ra \mu^\pm e^\mp$}

%{Short Distance Dominated Processes}

\bibitem{bigi}{I.~Bigi and F.~Gabbiani, {\it Nucl. Phys.} {\bf B367} 3 (1991).}

\bibitem{axion}{S.~Weinberg, {\it Phys.\ Rev.\ Lett.} {\bf 40} 223 (1978);
F.~Wilczek, {\it Phys.\ Rev.\ Lett.} {\bf 40} 279 (1978); T.~Goldman~and~C.M.~
Hoffman, {\it Phys.\ Rev.\ Lett.} {\bf 40} 220 (1978); J.M.~Fr\'ere, et al.,
{\it Phys.\ Lett.} {\bf B103} 129 (1981).}

\bibitem{familon}{F.~Wilczek, {\it Phys.\ Rev.\ Lett.} {\bf 49} 1549 (1982).}

\bibitem{hyper}{M.~Suzuki, {\it Phys.\ Rev.\ Lett.} {\bf 56} 1339 (1986);
S.H.~Aronson, et al., {\it Phys.\ Rev.\ Lett. } {\bf 56} 1342 (1986);
C.~Bouchiat~and~J.~Iliopoulos, {\it Phys.\ Lett. } {\bf 169B} 447 (1986).}

\bibitem{pnn787}{M.S. Atiya, et al., BNL48066, submitted to
{\it Phys.\ Rev.\ Lett.\ };
M.S. Atiya, et al., {\it Phys.\ Rev.\ Lett.\ } {\bf 64} 21 (1990).}

\bibitem{pnn2}{M.S. Atiya, et al., BNL48091, submitted to
{\it Phys.\ Rev.\ Lett.\ }.}

\bibitem{litt}{L.~Littenberg, {\it Phys. Rev.} {\bf D39} 3322 (1989).}

\bibitem{bunew}{G.~Buchalla and A.~Buras, MPI-PTh 2-93.}

\bibitem{pnn731}{G. E. Graham, {\it et al.}, {EFI-92-20} Apr. 1992.}

\bibitem{kami}{K. Arisaka, {\it et al.}, FNAL Jun. 1991.}

\bibitem{tini}{T. Inagaki,~T. Sato,~and~T. Shinkawa, {\it Experiment
to search for the Decay $K_L^0 \ra \pi^0 \nu \overline{\nu}$ at KEK
12 GeV Proton Synchrotron}, 30~November~1991.}

%{Radiative Decays}

%{$K_S \ra \gamma \gamma$}

\bibitem{epro}{G.~Ecker, A.~Pich and E.~de~Rafael, {\it Phys. Lett.}
{\bf 189B}  363 (1987);  L.~Cappiello and G.~D'Ambrosio,
{\it Nuov. Cim.} {\bf 99A} 153 (1988).}

\bibitem{goity}{G.~D'Ambrosio and D.~Espriu, {\it Phys. Lett.} {\bf 175B}
237 (1986); J.~Goity, {\it Zeit. Phys.} {\bf C34} 341 (1987).}

\bibitem{kssna}{H.~Burkhardt {\it et. al.}, {\it Phys. Lett.} {199B}
139 (1987).}

%{$K_L \ra \pi^0 \gamma \gamma$}

% refer to wdmw
\bibitem{bdv}{J.~Bijnens, S.~Dawson and G.~Valencia, {\it Phys. Rev.}
{\bf D44} 3555 (1991).}

\bibitem{pggNA31}{G.~D.~Barr, {\it et. al.}, {\it Phys. Lett.}
{\bf 242B} 523 (1990);G.~D.~Barr, {\it et, al.}, {\it Phys. Lett.}
{\bf 284B} 440 (1992).}

\bibitem{pgg731}{V.~Papadimitriou {\it et. al.}, {\it Phys. Rev.}
{\bf D44} R573 (1991).}

\bibitem{segg}{L.~M.~Sehgal, {\it Phys. Rev.} {\bf D41} 161 (1990);
P.~Ko, {\it Phys. Rev.} {\bf D41} 1531 (1990).}

\bibitem{lin} J.~Donoghue, B.~Holstein and Y.~C.~Lin, {\it Nucl. Phys.}
{\bf B277} 651 (1986).

\bibitem{cami}{L.~Cappiello, G.~D'Ambrosio, and M.~Miragliuolo,
{\it Phys. Lett.} {\bf 298B} 423 (1993).}

%{$K \ra \pi \gamma^*$}

%{$K^+ \ra \pi^+ \gamma \gamma$}

\bibitem{pgg787}{M.~S.~Atiya, {\it et. al.}, {\it Phys. Rev. Lett.}
{\bf 65} 1188 (1990).}

\bibitem{gilwise}{F.~Gilman and M.~Wise, {\it Phys. Rev.} {\bf D21}
3150 (1980).}

%{$K_L \ra \gamma \gamma$}

\bibitem{vene} G.~Shore and G.~Veneziano, {\it Nucl. Phys.} {\bf B381}
3 (1992).

%{$K_L \ra \gamma \gamma$ Dalitz decays}

\bibitem{eeg845}{K.E.~Ohl, et al., {\it Phys.\ Rev.\ Lett.\ } {\bf 65},
1407 (1990).}

\bibitem{eegNA31}{G.~D.~Barr, {\it et. al.}, {\it Phys. Lett.}
{\bf 240B} 283 (1990)}

\bibitem{mdexp}{R. Tschirhart, talk at DPF Meeting at FNAL, Nov. 1992.}

\bibitem{berg}{L.~Bergstr\"{o}m, E.~Mass\'{o}, P.~Singer,
{\it Phys. Lett.} {\bf 131B} 229 (1983);
L.~Bergstr\"{o}m, {\it et. al.},
{\it Phys. Lett.} {\bf 134B} 373 (1984).}

\bibitem{quijak}{C.~Quigg and J.~D.~Jackson, UCRL-18487 unpublished (1968).}

\bibitem{krollw}{N.~Kroll and W.~Wada, {\it Phys. Rev.} {\bf 98}
1355 (1955).}

\bibitem{dedexp}{G.~D.~Barr, {\it et. al.}, {\it Phys. Lett.} {\bf 259B}
389 (1991).}

\bibitem{eeee845}{M.R.~Vagins, {\it et. al.}, YAUG-A-93/1, Jan. 1993,
submitted to {\it Phys. Rev. Lett.} }

\bibitem{eeee137}{T. Akagi, et al., {\it KEK Preprint 92-35,\ } May 1992.}

\bibitem{eeee799}{Ping Gu, talk at DPF Meeting at FNAL, Nov. 1992.}

\bibitem{miyaz}{T.~Miyazaki and E.~Takasugi, {\it Phys. Rev.} {\bf D8}
2051 (1973).}

%{Direct emission $K_L \ra \pi^+ \pi^- \gamma$}

\bibitem{oldl}{C.~S.~Lai, and B.~L.~Young, {\it Nuov. Cim.} {\bf 52A}
83 (1967);
L.~Sehgal and L.~Wolfenstein, {\it Phys. Rev.} {\bf 162} 1362 (1967);
D.~Beder, {\it Nucl. Phys.} {\bf B47} 286 (1972).}

\bibitem{coska}{G.~Costa and P.~Kabir, {\it Il Nuov. Cim.} {\bf LI A}
N.2 564 (1967).}

\bibitem{carroll}{A.~S.~Carroll, {\it et. al.}, {\it Phys. Rev. Lett.}
{\bf 44} 529 (1980); E{\bf 44} 1026 (1980).}

\bibitem{kppgfnal}{G.~J.~Bock, {\it et. al.}, FNAL 92/384-E;E. Ramberg,
 {\it et al.}, FNAL 92/385-E}

\bibitem{linv}{Y.~C.~Lin and G.~Valencia, {\it Phys. Rev.} {\bf D37}
143 (1988).}

\bibitem{piccio}{C.~Picciotto, {\it Phys. Rev.} {\bf D45} 1569 (1992).}

\bibitem{enp}{G.~Ecker, H.~Neufeld and A.~Pich, {\it Phys. Lett.}
{\bf 278B} 337 (1992).}

\bibitem{manf}{L.~Sehgal and M.~Wanninger, {\it Phys. Rev.} {\bf D46}
1035 (1992); E {\bf D46} 5209 (1992).}

%{Direct Emission in $K^+ \ra \pi^+ \pi^0 \gamma$}

%refer to enp

%{$K_L \ra \ell^+ \ell^-$}

%{Short Distance}

\bibitem{herczeg}{P.~Herczeg, {\it Phys. Rev.} {\bf D27} 1512 (1983).}

\bibitem{epmupo}{G.~Ecker and A.~Pich, {\it Nucl. Phys.} {\bf B366}
189 (1991).}

\bibitem{botlim}{F.~Botella and C.~S.~Lim, {\it Phys. Rev. Lett.} {\bf 56}
1651 (1986).}

%{Long distance}

\bibitem{mumu137}{T. Akagi, et al., {\it Phys.\ Rev.\ Lett.\ } {\bf 67}
2618 (1991).}

\bibitem{ekmm}{A.~J.~Schwartz, Princeton/hep/92-15.}

\bibitem{simipi}{M.~Savage, M.~Luke and M.~Wise, {\it Phys. Lett.}
{\bf 291B} 481 (1992).}

\bibitem{quigg}{C.~Quigg, private communication.}

\bibitem{belgeng}{G.~B\'{e}langer and C.~Q.~Geng, {\it Phys. Rev.}
{\bf D43} 140 (1991); P.~Ko, {\it Phys. Rev.} {\bf D45} 174 (1992).}

%{$K\ra \pi \ell^+ \ell^-$}

%{$K^+ \ra \pi^+ \ell^+ \ell^-$}

\bibitem{alliegro}{C.~Alliegro {\it et. al}, {\it Phys. Rev. Lett.} {\bf 68}
278 (1992).}

\bibitem{pmm787}{M.~S.~Atiya, {\it et. al.}, {\it Phys. Rev. Lett.}
{\bf 63} 2177 (1989).}

\bibitem{savawi}{M.~Savage and M.~Wise, {\it Phys. Lett.} {\bf 250B}
151 (1990).}

\bibitem{sawi}{M.~Lu, M.~Wise and M.~Savage, {\it Phys. Rev.} {\bf D46}
5026 (1992).}

\bibitem{bg}{G.~B\'{e}langer, C.~Q.~Geng and P.~Turcotte, UdeM-LPN-TH-90.}

\bibitem{beng}{P.~Agrawal {\it et. al.}, {\it Phys. Rev. Lett.} {\bf 67}
537 (1991); {\it Phys. Rev.} {\bf D45} 2383 (1992).}

%{$K_1^0 \ra \pi^0 \ell^+ \ell^-$}

%{$K_L \rightarrow \pi^0 e^+ e^-$}

\bibitem{otherpiee}{C. Dib, I. Dunietz, and F.J. Gilman, {\it Phys. Rev.}
{\bf D39} 2639 (1989);
C. Dib, I. Dunietz, and F.J. Gilman  {\it Phys. Lett.} {\bf 218B} 487 (1989);
J.M. Flynn, {\it Nucl. Phys.} {\bf B13} 474 (1990);
J.M. Flynn and L. Randall, {\it Nucl. Phys.} {\bf B326} 31 (1989),
E.-{\it ibid.} {\bf B334} 580 (1990).}

\bibitem{seg}{J.~Donoghue, B.~Holstein and G.~Valencia, {\it Phys. Rev.}
{\bf D35} 2769 (1987);
L.~M.~Sehgal, {\it Phys. Rev.} {\bf D38} 808 (1988);
J.~Flynn and L.~Randall, {\it Phys. Lett.} {\bf 216B} 221 (1989).}

\bibitem{peeNA31}{G.D. Barr, et al., {\it Phys.\  Lett.\ } {\bf B214} 303
(1988).}

\bibitem{pee731}{A. Barker, et al., {\it Phys.\ Rev.\ } {\bf D41} 3546 (1990).}


\bibitem{kpee845}{K.E.~Ohl, et al., {\it Phys.\ Rev.\ Lett.\ } {\bf 64},
2755 (1990).}

\bibitem{eegg845}{W.M.~Morse, et al., {\it Phys.\ Rev.\ } {\bf D65},
36 (1992).}

\bibitem{green}{H.B.~Greenlee, {\it Phys.\ Rev.\ } {\bf D42} 3724 (1992).}

%{$K_L \ra \pi^0 \mu^+ \mu^-$}

%{Experiments}

%\bibitem{me}{L.S.~Littenberg,``CP violation in $K_L \to \pi^0 e^+ e^-$'',
%{\sl Proceedings of the Workshop on CP Violation at KAON Factory},
%December 3, 1988, ed. J.N. Ng, p. 19.}

\bibitem{e162}{K. Miyake, et al.,{\sl Measurement of CP-violating direct
amplitude in $K_L \to \pi^0 e^+ e^-$ decay}, Aug. 1988.}

\bibitem{Spee731}{L. K. Gibbons, et al., {\it Phys.\ Rev.\ Lett.\ } {\bf 61}
2661 (1988).}

\bibitem{lightNA31}{G.~D.~Barr, {\it et. al.}, {\it Phys. Lett.}
{\bf 235B} 356 (1990)}

\bibitem{H777}{N.J.~Baker, et al., {\it Phys.\ Rev.\ Lett.\ } {\bf 59}
2832 (1987).}

\bibitem{mumu791}{A. Schwartz., in {\sl Fourth Conference on the Intersections
between Particle and Nuclear Physics}, Tucson, Arizona, AIP Conference
proceedings No 243, ed. W. T. H. Van Oers (AIP, New York), p. 609.}

\bibitem{app787}{M.S. Atiya, et al., {\it NIM. } {\bf A321} 129 (1992).}

\end{thebibliography}
\end{document}